\documentclass[12pt]{article}

\usepackage{epsfig}
\usepackage{graphicx}
\usepackage{mathtools}
\usepackage{commath}
\usepackage{booktabs}
\usepackage{amsmath}
\usepackage{amssymb}
\usepackage{latexsym}
\usepackage{color}
\usepackage[T1]{fontenc}
\usepackage{authblk}
\usepackage{mathrsfs}
\usepackage{natbib}
\usepackage{hyperref}
\usepackage{appendix}
\usepackage[left=2.5cm,top=4cm,right=2.5cm,bottom=3cm]{geometry}
\usepackage{subfig}
\usepackage{float}
\usepackage{caption}
\usepackage{caption}
\usepackage{xcolor}
\usepackage{comment}
\providecommand{\keywords}[1]
{\small\textbf{\textit{Keywords--}} #1}

\title{Feature detection 
in point processes on linear networks using nearest neighbour volumes}
\author[1]{Juan F. D\'iaz-Sep\'ulveda\thanks{Corresponding author: jfdiazs0@unal.edu.co (Juan F. D\'iaz-Sep\'ulveda).}}
\author[2]{Nicoletta D'Angelo}
\author[2]{Giada Adelfio}
\author[3]{Jonatan A. Gonz\'alez}
\author[1]{Francisco J. Rodr\'iguez-Cort\'es}
\affil[1]{\small Escuela de Estad\'istica, Universidad Nacional de Colombia, Medell\'in, Colombia}
\affil[2]{\small Dipartimento di Scienze Economiche, Aziendali e Statistiche, Universit\`a degli Studi di Palermo, Palermo, Italy}
\affil[3]{\small Department of Mathematics, University Jaume I, Castellon, Spain. jmonsalv@uji.es}
\date{\date{}}

\begin{document}
\maketitle

\begin{abstract}
We consider the feature detection problem in the presence of clutter in point processes on linear networks. We extend the classification method developed in previous studies to this more complex geometric context, where the classical properties of a point process change and data visualization are not intuitive. We use the $K$-th nearest neighbour volumes distribution in linear networks for this approach. As a result, our method is suitable for analysing point patterns consisting of features and clutter as two superimposed Poisson processes on the same linear network. To illustrate the method, we present simulations and examples of road traffic accidents that resulted in injuries or deaths in two cities in Colombia.

\end{abstract}

\section{Introduction}
In the last decade, spatial statistics has undergone an extraordinary methodological and computational advancement focused on generalising and extending the foundations of the theories to more complex geometric spaces that allow fairer statistical analysis of new kinds of spatial data. In particular, spatial point process methodologies have been addressed to consider non-classical geometric supports for analysing events on linear networks, such as traffic accidents in a city \citep{BADDELEY2021100435}. One of the most relevant contributions to the statistical analysis of network data is the definition of the geometrically corrected first- and second-order characteristics. These descriptors can be used to identify spatial configurations and discriminate between point patterns according to their aggregation structure \citep{Ang2012,MBN17,MRCM18,RBN19,RDMMGMB19,dangelo2021assessing,dangelo2022spst}, select or reduce a set of variables \citep{RMcNB21} and estimate relative risks \citep{MBN2020}.

A frequent research topic for spatial data analysis  is identifying features of events in the presence of clutter. The conventional terminology is that a \textit{feature} is a point of the pattern or process of interest, and \textit{clutter} consists of extraneous points that are not proper to the pattern of interest. For instance, detecting surface minefields from an image from a reconnaissance aircraft can be processed to obtain a list of objects, some of which may be mines and others any other type of object \citep{1997-AF, Byers1998}. For spatial point processes, this problem has been addressed differently, either denoted by \textit{feature detection} or \textit{clutter removal}.  \cite{1997-AF} developed a method to find the maximum likelihood solution using Voronoi polygons. \cite{dasgupta1998detecting} used model-based clustering to extend the methodology proposed by \cite{banfield1993model}.  While these methods are based on some limiting assumptions, \cite{Byers1998} adopted a different approach in which they estimated and removed the clutter without making any assumptions about the shape or number of features. More recently, \cite{gonzalez2021classification} considers the local contributions of the pair correlation function as functional data and two classification procedures to separate features from clutter points are described.

However, when minefields are distributed over the streets of a city or a neighbourhood, i.e., over a linear network, this problem cannot be analysed similarly because the events do not occur in the whole plane.  Indeed, \cite{BADDELEY2021100435} states that statistical analysis of point data on linear networks presents severe problems because the associated graph  is not spatially homogeneous, creating geometric and computational complexities and leading to new problems with a high risk of methodological error. In addition, data on networks can have a wide range of spatial scales. \cite{BADDELEY2021100435} emphasise that the point processes on linear networks also challenge the classical methodology of spatial statistics based on stationary processes, which is mainly inapplicable to network data. For these reasons, in this paper, we extend the spatial technique proposed by \cite{Byers1998} from the planar case to the linear network one. Following \cite{Byers1998}, we use the observed nearest neighbour volumes defined by the $K$-th nearest neighbour distances modelled as a mixture distribution and estimate the parameters by an $EM$ algorithm to classify data points as features or otherwise. 

The structure of the paper is as follows. Section \ref{sec:framework} gives some basic techniques for analysing point patterns on linear networks. Section \ref{sec:approach} presents the proposed method for feature detection on linear networks. Section \ref{sec:simulations} shows an extended simulation study carried out with different linear networks. Section \ref{sec:bogtraffic} contains two applications of the classification method for traffic accidents in Bogota and Medellin (Colombia). Section \ref{sec:conclusions} presents some conclusions. 

\section{Mathematical framework}\label{sec:framework}
Following~\citep{Ang2012,BADDELEY2021100435}, we consider a linear network as the union of a finite number of line segments on the plane, namely  $L=\bigcup_{i=1}^n l_i$, such that each $l_i = [u_i,v_i]=\{w:w=t u_i+(1-t)v_i, 0\leq t \leq 1\}$, and $u_i$, $v_i$ are the endpoints of the segment $l_i$. The volume of the network $L$, denoted by $\abs{L}$, is the total length of the segments contained in thereof. Additionally, a path between two points $u$ and $v$ in a linear network $L$ is a sequence $x_0,x_1,\hdots,x_m$ of points in $L$ so that $x_0=u$, $x_m=v$ and $[x_i, x_{i+1}] \subset L$ for each $i = 0,\hdots,m-1$. The path length is the sum of the lengths of its segments. The shortest path distance $d_L(u,v)$ between $u$ and $v$ in $L$ is the minimum of the lengths of all paths from $u$ to $v$. If there are no paths from $u$ to $v$ (implying that the network is not connected), then the distance is defined by $d_L(u,v)=\infty$.

We also assume that the disc of radius $r>0$ and centre point $u\in L$ is the set of all points $v$ in the network lying no more than a distance $r$ from $u$, that is $b_L(u,r)=\{v\in L:d_L(u,v)\leq r\}$. The relative boundary of the disc is the set of points lying exactly $r$ units away from $u$, $\partial b_L(u,r)=\{v\in L: d_L(u,v)=r\}$. The circumference $m(u, r)$ is the number of points of $L$ that fall in $\partial b_L(u,r)$. The circumference is finite for all $r<\infty$, and we set $m(u,\infty)=\infty$ by convention. The circumradius of the network is the radius of the smallest disc that contains the entire network, $R=R(L)=\inf\{r:m(u,r)=0,\mbox{ for some $u \in L$}\}={\min}_{u \in L} \ {\max}_{v \in L} \ d_L(u,v)$. If $L$ is not path-connected, then $R$ is the minimum of the circumradius of the connected components of $L$; for more details see~\citep{Ang2012,BADDELEY2021100435}.

We define a (finite, simple) point pattern $X$ on $L$ as a ﬁnite set $X=\{x_1,x_2,\ldots,x_n\}$ of distinct points $x_i\in L$, where $n \geq 0$ \citep[see e.g.,\ ][]{RMcNB21}. Further, $N_X(B)=N(X\cap B)$ is the number of points on $X$ lying in $B$ for any set $B\subset L$. Thus, for a ﬁnite and simple point possess with intensity function $\lambda(u)$, $u\in L$, we have $\mathbb{E}[N_X(B)]=\Lambda(B)=\int_{B}\lambda(u)\mbox{d}_1 u$, for all measurable $B\subset L$, where $\mbox{d}_1 u$ denotes integration with respect to arc length~\citep{federer2014geometric}. If $\lambda$ is constant, then $X$ is called homogeneous and the constant $\lambda$ is the mean number of points per unit length for a homogeneous point process. A homogeneous Poisson point process on $L$ with intensity $\lambda > 0$ is characterized by the properties that, for any line segment $B \subset L$, the number of points falling in $B$ has a Poisson distribution with mean $\lambda \abs B$, while events occurring in disjoint line segments $B_1,\hdots,B_m \subset L$ are independent \citep{Ang2012}. Finally, a sub-network of $L$ is a linear network, a subset of $L$.

\section{Approach for clutter removal on linear networks}\label{sec:approach}
In spatial point pattern analysis, some authors have been interested in identifying features surrounded by clutter. Unlike the planar case, which is very rarely straightforward, identifying network features through visual inspection is more difficult due to the complexity of the domain. We consider the extension of the approach in \cite{Byers1998} and assume that the clutter is distributed as a homogeneous Poisson point process on a linear network with some intensity $\lambda_c$. The features are also homogeneous Poisson point processes with an equal or different intensity $\lambda_f$ than clutter, restricted to a certain sub-network and overlaid on the clutter. Therefore, the resulting point process is Poisson with piece-wise constant intensity on the linear network.
\begin{figure}[!ht]
\centering
\subfloat[Feature sub-network of \texttt{chicago}.]{\includegraphics[width=0.3\textwidth]{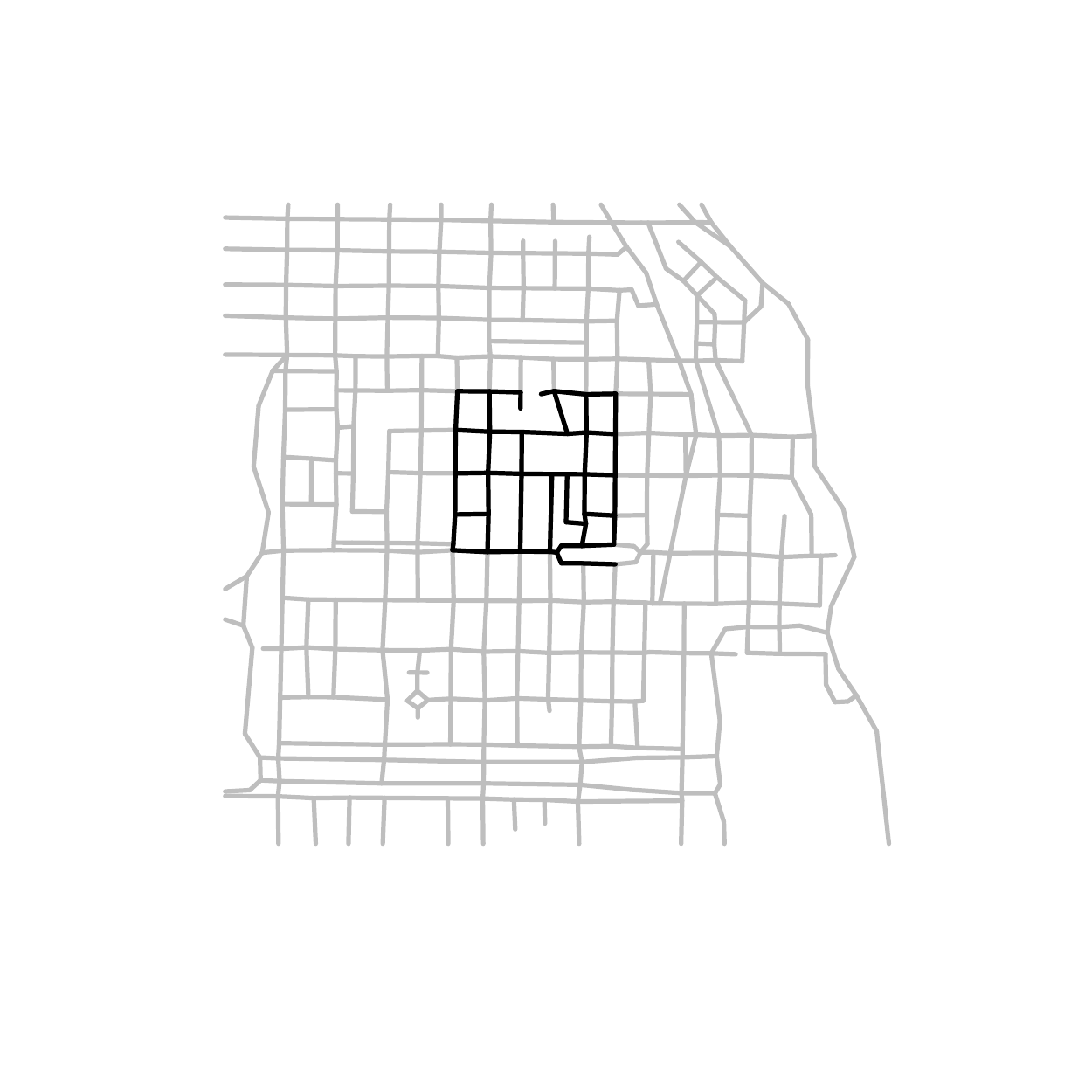}\label{fig:chicago-a}}\qquad\qquad
\subfloat[Clutter point pattern.]{\includegraphics[width=0.3\textwidth]{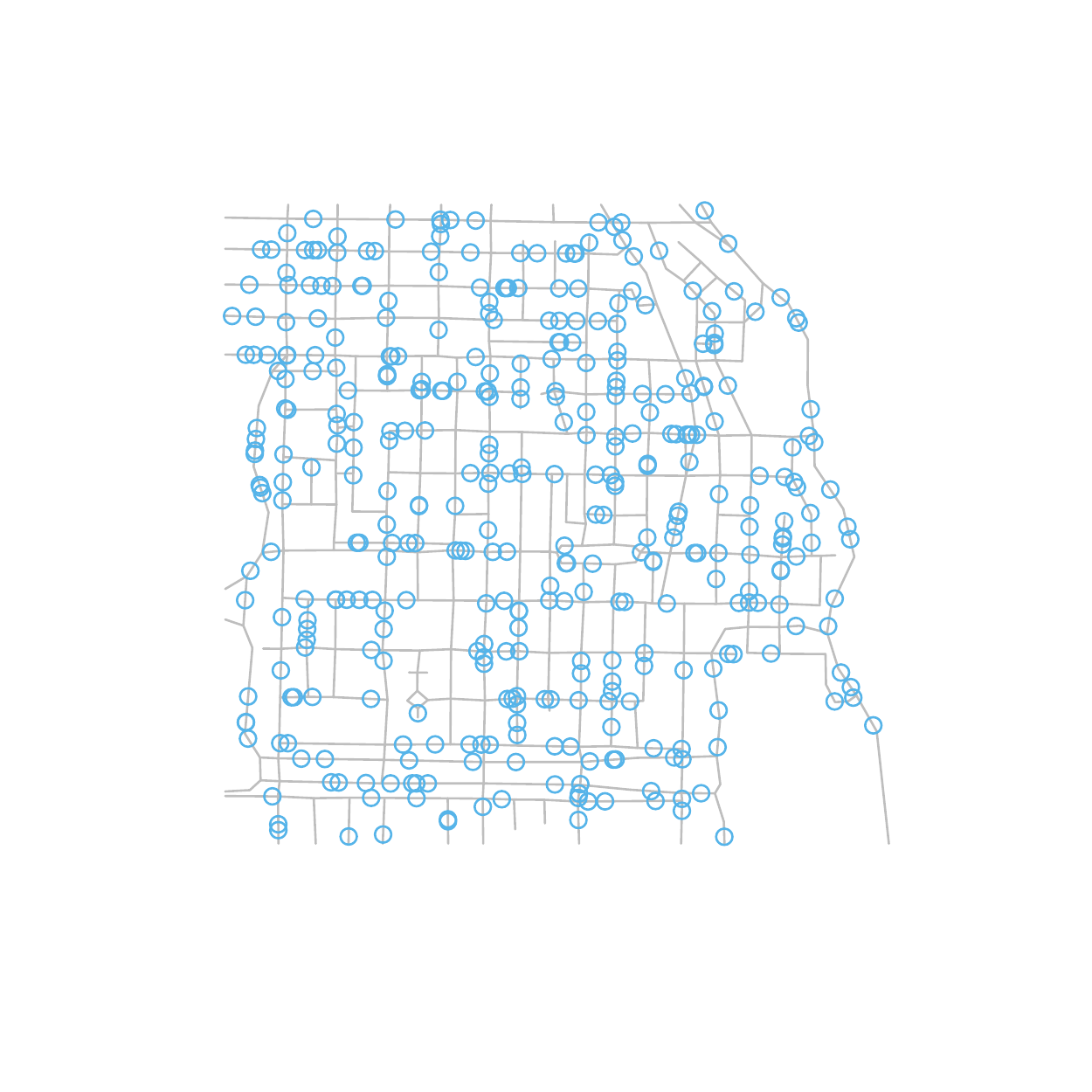}\label{fig:chicago-b}}\\
\subfloat[Feature point pattern.]{\includegraphics[width=0.3\textwidth]{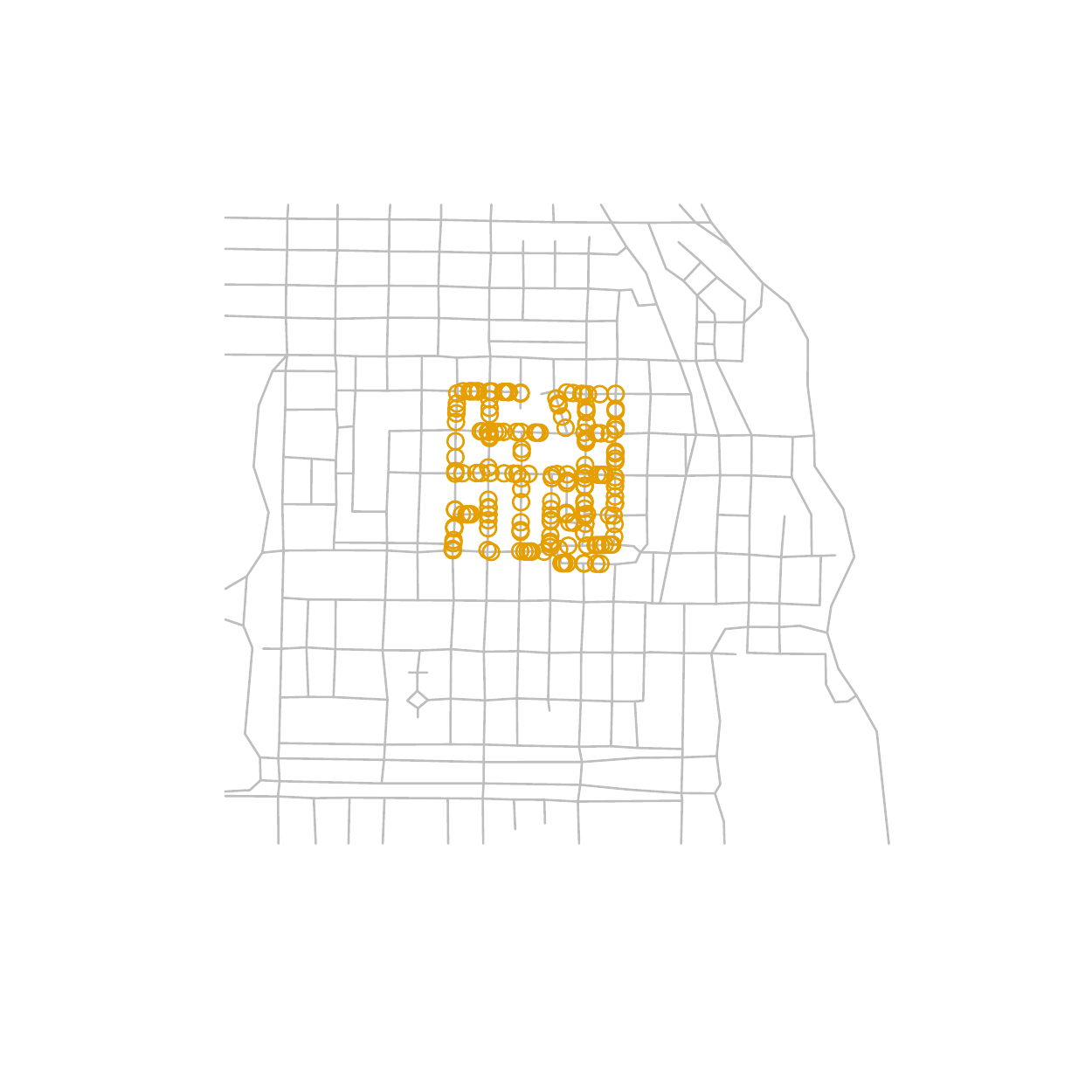}\label{fig:chicago-c}}\qquad\qquad
\subfloat[Superposition of clutter and feature. ]{\includegraphics[width=0.3\textwidth]{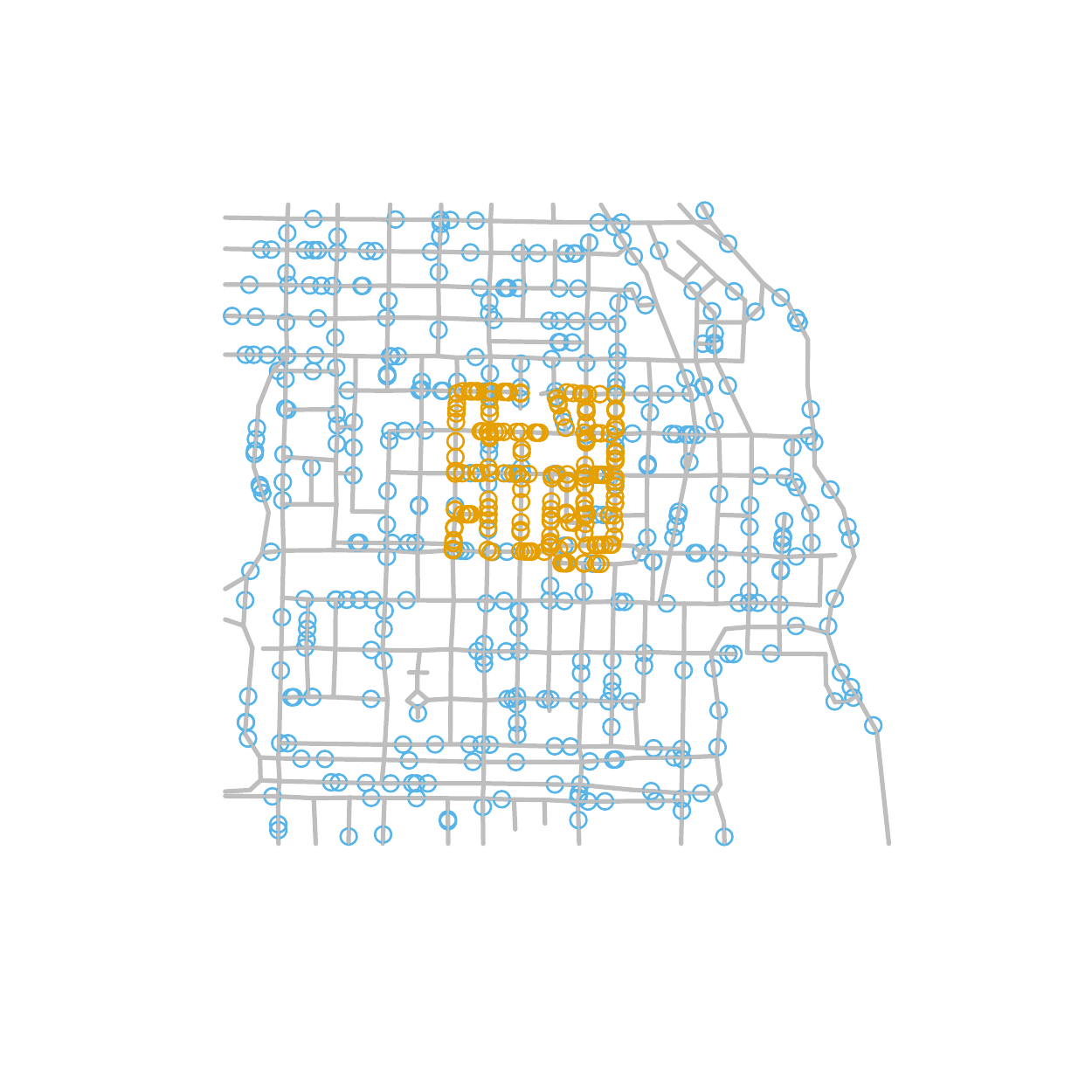}\label{fig:chicago-d}}
\caption{Linear network \texttt{chicago} at four different stages. A clutter point pattern simulated from a homogeneous Poisson point process with $\lambda_{c}=0.013$ and $\mathbb{E}[n_c]=400$. The feature is an observation of a homogeneous Poisson point process with $\lambda_{f}=0.067$ and $\mathbb{E}[n_f]=200$.}
\label{fig:chicago}
\end{figure}

In order to illustrate the geometric context for our developments, we consider the linear network \texttt{chicago} of the package \texttt{spatstat.data} \citep{spatstat.data}, which represents the street network of some area of Chicago (Illinois, USA) close to the University of Chicago; a particular case of simulated data on this linear network is shown in Figure~\ref{fig:chicago} at four different stages \citep{Ang2012}. The sub-network where the features are generated is shown in Figure~\ref{fig:chicago-a}. The street network has 338 intersections, 503 uninterrupted segments and a total length of 31150 feet. The sub-network has 39 intersections, 53 uninterrupted segments and a total length of 2991 feet. Figure~\ref{fig:chicago-b} shows the clutter point pattern simulated from a homogeneous Poisson point process with intensity $\lambda_{c}=0.013$ and expected number of points $\mathbb{E}[n_c]=400$ across the entire linear network. Figure~\ref{fig:chicago-c} shows the feature point pattern simulated from a Poisson process with intensity $\lambda_{f}=0.067$ and expected number of points $\mathbb{E}[n_f]=200$ in the sub-network. The superposition of the clutter and feature point patterns in Figures~\ref{fig:chicago-b} and~\ref{fig:chicago-c} are displayed in Figure~\ref{fig:chicago-d}, which is the resulting point pattern, that is, what we typically observe.

\subsection{\textit{K}-th nearest neighbour volumes distribution}

If the number of independent random events occurring in a region of a specific area is a Poisson variable with constant intensity $\lambda$, then for an integer $\alpha$, the required area for the occurrence of $\alpha$-th Poisson events in a sub-region has a Gamma distribution, $\Gamma(\alpha, \lambda)$, where shape and rate parameters are $\alpha$ and $\lambda$, respectively \citep{canavos1984applied}. In the same way, it can be said that if the number of independent random events occurring in a linear network is a Poisson variable with constant intensity $\lambda$, the sub-network volume that is necessary for $\alpha$-th Poisson events to fall within this sub-network has a distribution $\Gamma(\alpha, \lambda)$.

In the linear networks geometry, the disc $b_L(u,r)$ has variable volume for a given $r$ and different locations $u$ (i.e. the disc volume depends on $u$), as mentioned in \cite{cronie2020}. Assuming a Poisson process $X$ on $L$ with constant intensity $\lambda$, the number of points $N_X(b_L(u,r))$ of $X$ falling in the disc $b_L(u,r)$, has an approximate Poisson distribution with mean equal to $\lambda|b_L(u,r)|$. Because $N_X(b_L(u,r))$ is truncated above by $|L|$, the Poisson distribution is an approximation.
For $u \in L$, let $D_K^{L}(u)$ be the distance of the $K$-th nearest neighbour of $u$ and $S_K(u)=|b_L(u,D_K^{L}(u))|$. As $N_X(b_L(u,D_K^{L}(u)))=K$, then $S_K=S_K(u)$ has an approximate Gamma distribution $ \Gamma(K, \lambda)$.

Accordingly, the density of \textit{K}-th nearest neighbour volume $S_K$ is
\begin{align}
f_{S_K}(x)&=\frac{\lambda^{K} \ x^{K-1} \ e^{-\lambda x}}{\Gamma(K)}, \ \ \mbox{for} \ x>0; \ K, \lambda>0.
\label{eq:density}
\end{align}
Having a closed-form and the Gamma distribution properties, the maximum likelihood estimation of the rate  given the observed values of $S_K$ is straightforward. Indeed, the maximum likelihood estimate of $\lambda$ is
\begin{equation}
    \hat{\lambda}=\frac{nK}{\sum^n_{i=1}s_i}=\frac{K}{\bar s_i},
    \label{eq:lambda}
\end{equation}
where $s_i=|b_L(x_i,D_K^{L}(x_i))|$ with $x_1,\hdots,x_n$ are points of the Poisson process $X$.

\subsection{Mixture modeling and estimation}
\begin{figure}[h!]
\centering
\includegraphics[width=.28\textwidth]{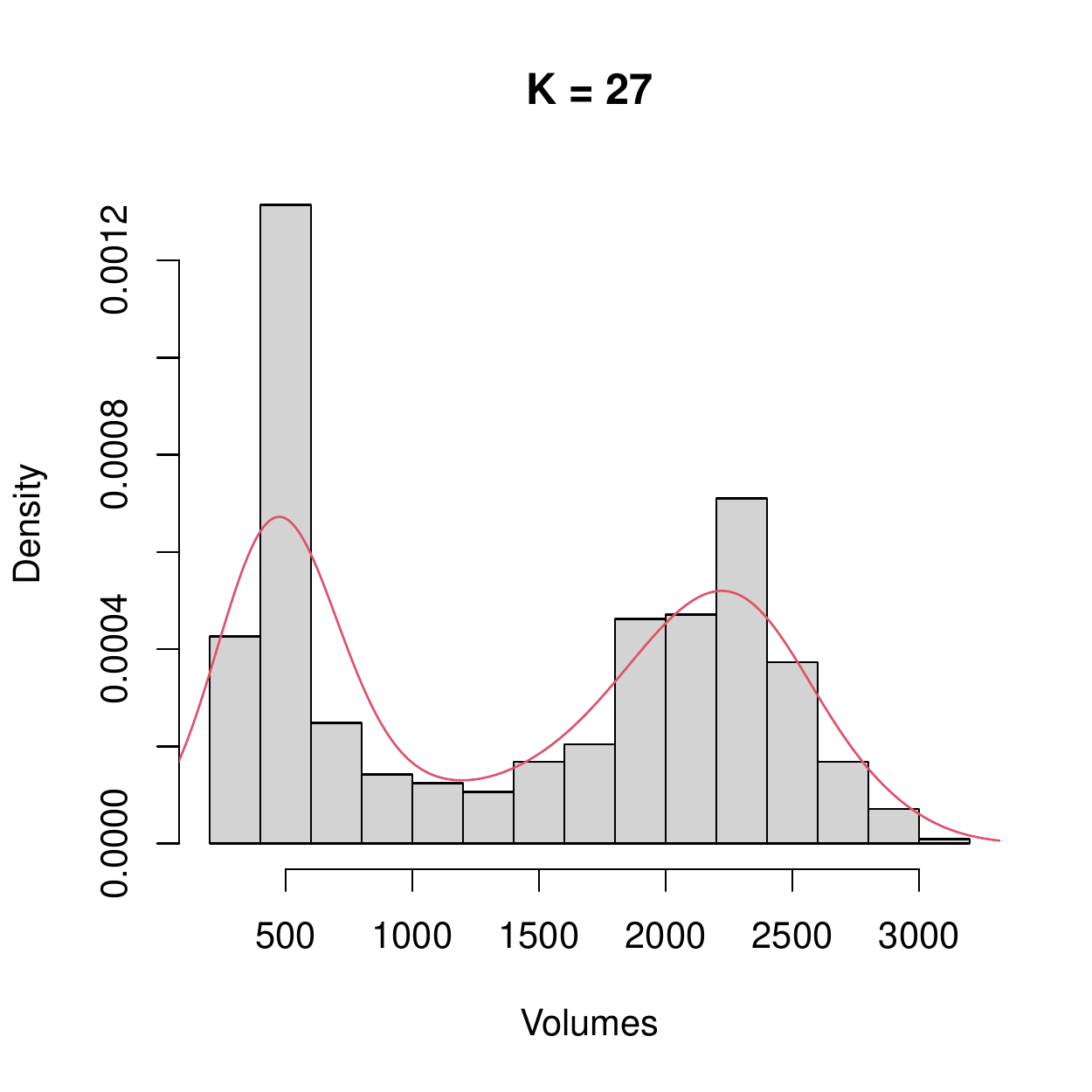}
\includegraphics[width=.28\textwidth]{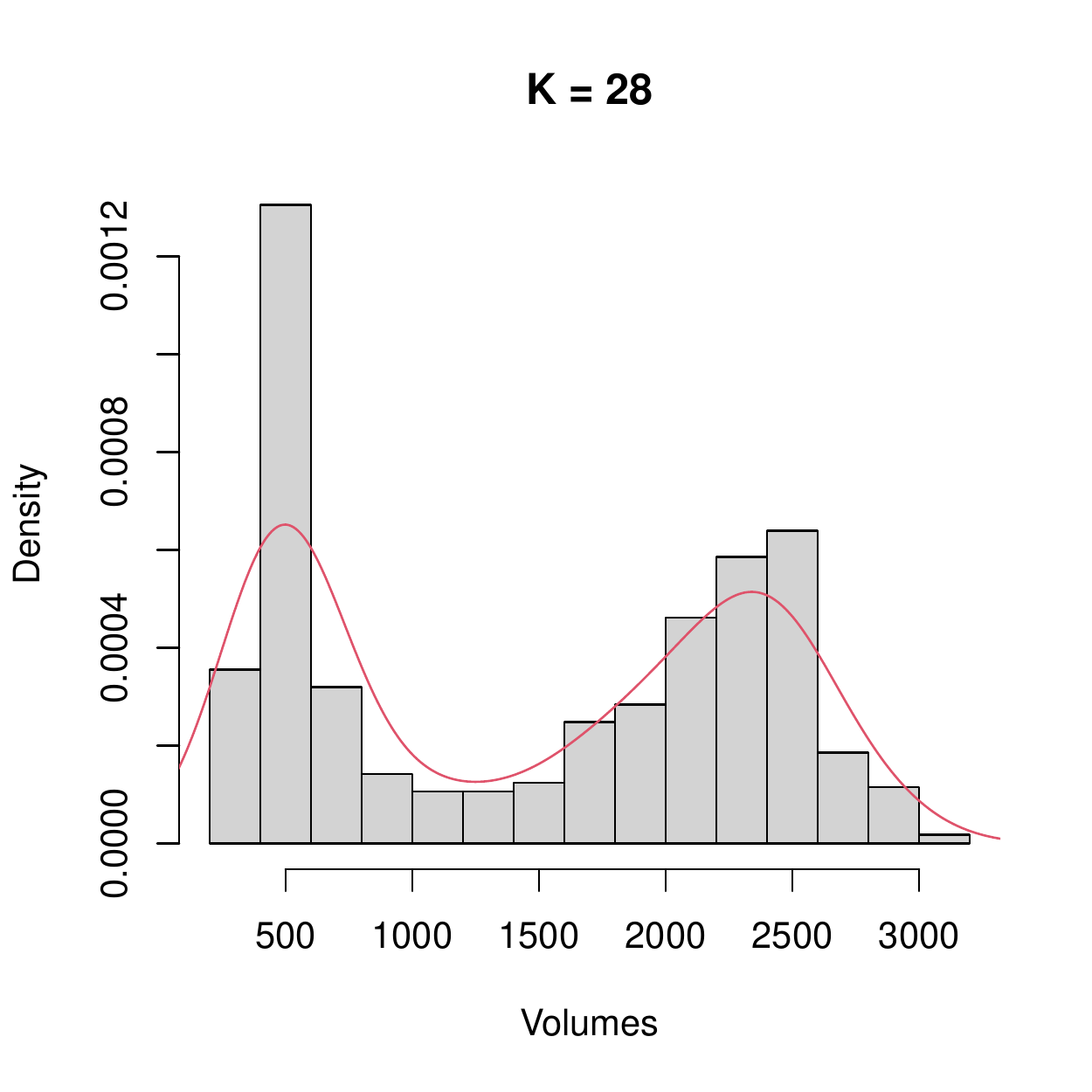}
\includegraphics[width=.28\textwidth]{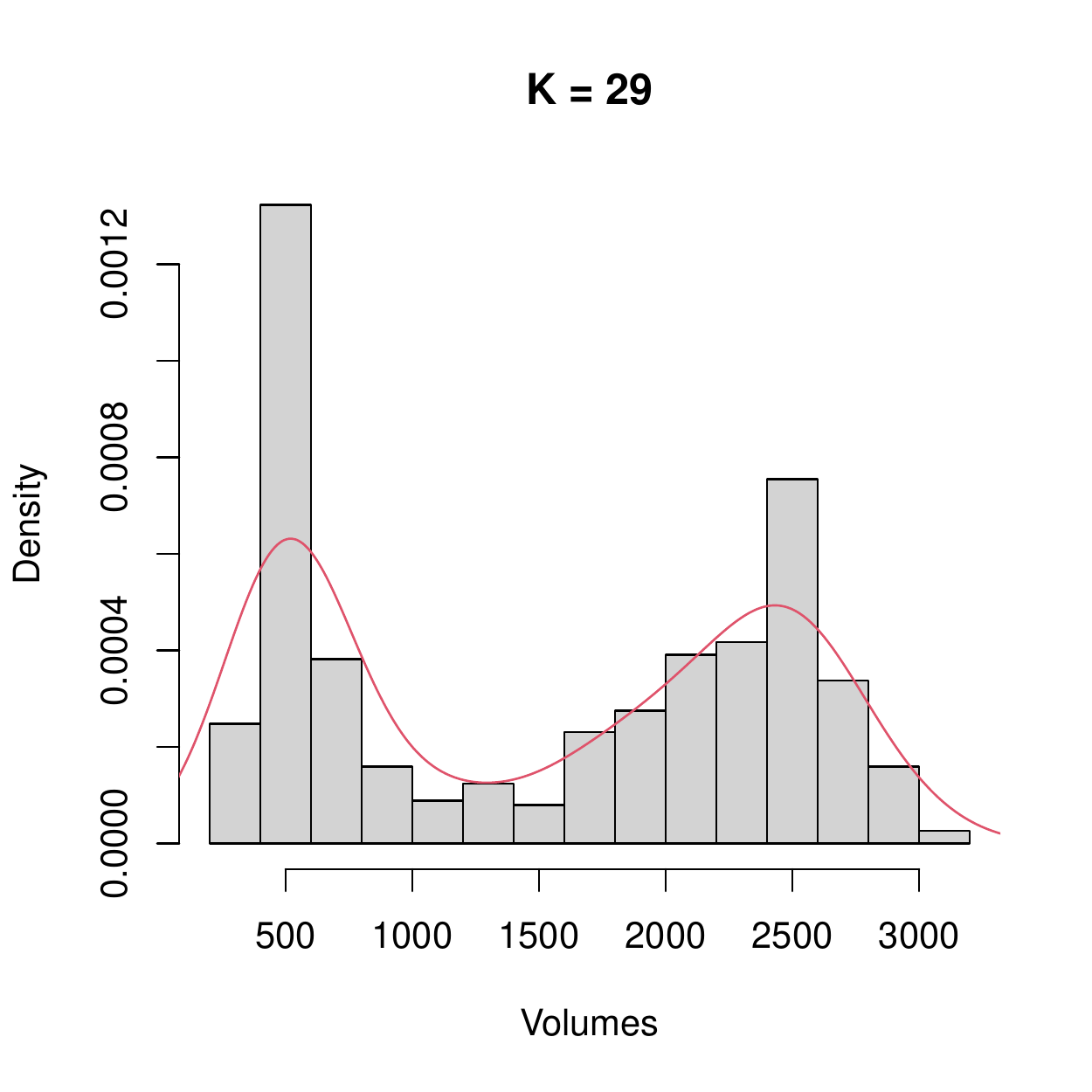}\\\vspace{0.4cm}
\includegraphics[width=.28\textwidth]{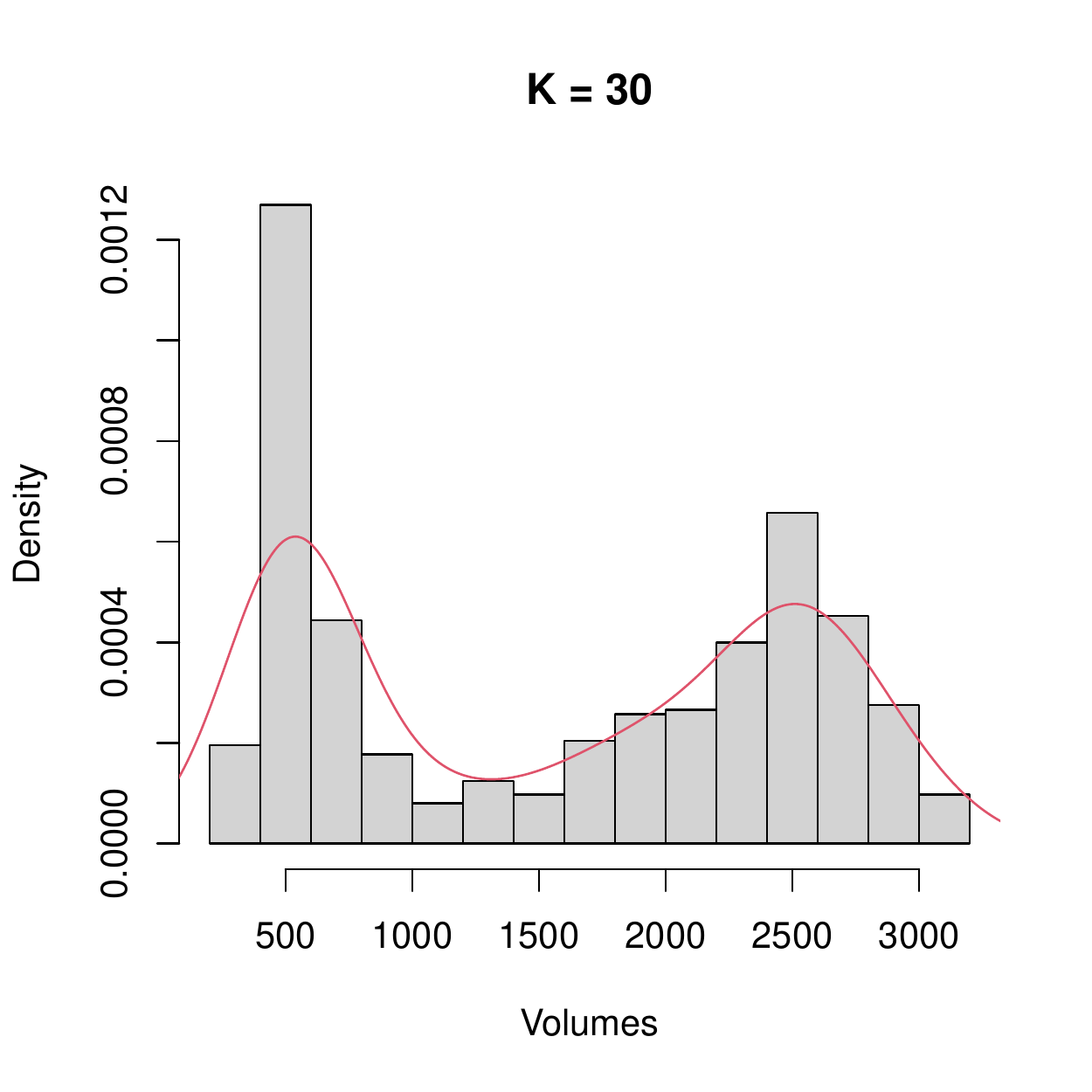}
\includegraphics[width=.28\textwidth]{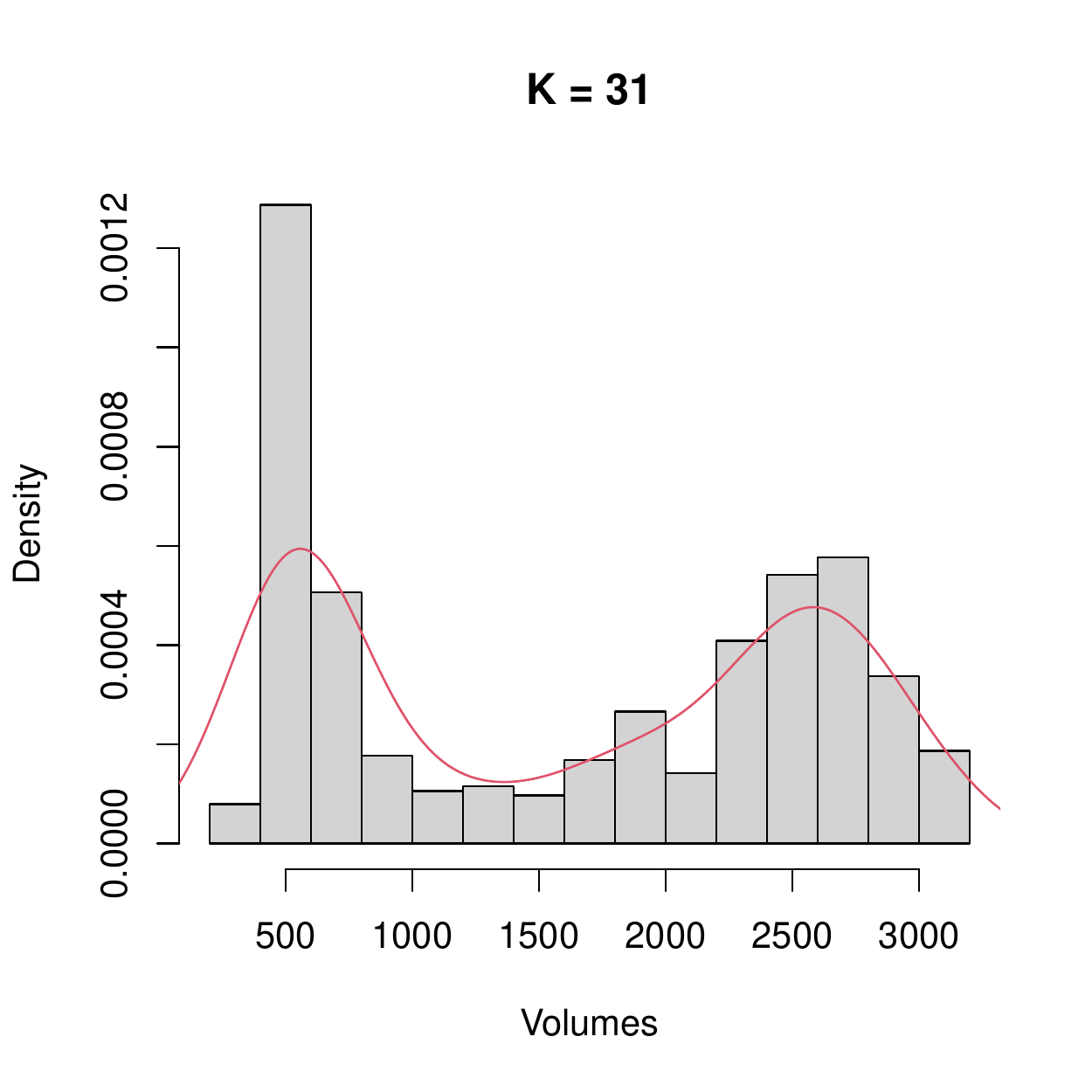}
\includegraphics[width=.28\textwidth]{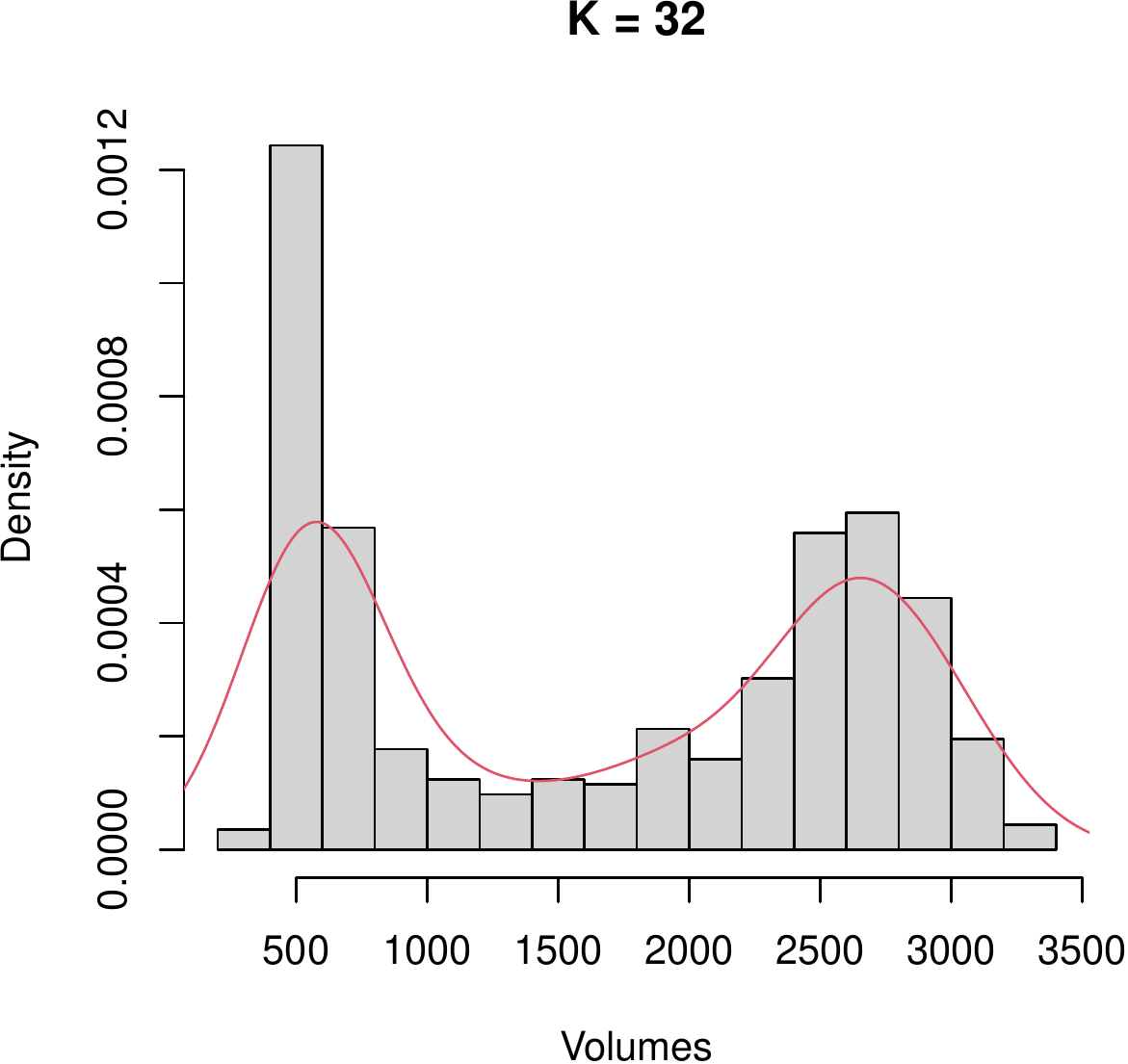}
\caption{\label{fig:Hist_nn27-30} Distributions of the volumes $S_K$ for the mixture of clutter and feature from $27$-th to $32$-nd nearest neighbours simulated on the \texttt{chicago} network in Figure~\ref{fig:chicago}.}
\end{figure}

Six estimated volume distributions $S_K$ for the mixture of a couple of homogeneous Poisson point patterns simulated in the \texttt{chicago} network (see Figure~\ref{fig:chicago}) from the $27$-th to the $32$-nd nearest neighbour are displayed as histograms in Figure~\ref{fig:Hist_nn27-30}, where a clear bimodality is evident. They can be used as exploratory tools and motivation to propose a model \citep{Byers1998}. We assume two types of processes to be classified through a mixture of the corresponding $K$-th nearest neighbour volumes coming from the clutter and feature, two superimposed Poisson processes. Therefore, based on equation \eqref{eq:density}, we assume that
\begin{equation}\label{mix-dist}
S_K \sim p \Gamma(K, \lambda_1)+(1-p) \Gamma(K, \lambda_2),
\end{equation}
where $\lambda_1$ and $\lambda_2$ are the intensities of the two homogeneous Poisson point processes (feature and clutter, respectively) on the linear network, and $p$ is the constant that characterizes the postulated distribution of the volumes $S_K$. Note that, intuitively, we assume feature points to be part of the first component, as being the ones with shorter distances from each other.

The parameters $\lambda_1, \lambda_2$ and $p$ associated with the mixture are estimated using an $EM$ algorithm \citep{DLR1977}, wherein we use the closed-form of a Gamma distribution in the expectation step. On the other hand, let $\delta_i \in \{0, 1\}$ be the two classification components for each data point, where $\delta_i= 1$ if the $i$-th point belongs to the feature and $\delta_i= 0$ otherwise. Thus each data point has an observed volume $s_i$ of $S_K$ and an unknown classification component $\delta_i$. Hence, the $E$ step of the algorithm consists of
\begin{equation}
\mathbb{E}[\hat \delta_i^{(t+1)}]=\frac{\hat p^{(t)}f_{S_K}(s_i;\hat \lambda_1^{(t)})}{\hat p^{(t)}f_{S_K}(s_i;\hat \lambda_1^{(t)})+(1-\hat p^{(t)})f_{S_K}(s_i;\hat \lambda_2^{(t)})},
\label{eq:stepE}
\end{equation}
and the maximization $M$ step consists of
\begin{equation}
\hat \lambda_1^{(t+1)}=\frac{K \sum_{i=1}^n\hat \delta_i^{(t+1)}}{\sum_{i=1}^n s_i\hat \delta_i^{(t+1)}}, \quad  \hat\lambda_2^{(t+1)}=\frac{K \sum_{i=1}^n(1-\hat \delta_i^{(t+1)})}{\sum_{i=1}^n s_i(1-\hat \delta_i^{(t+1)})} \quad \text{and}\quad \hat p^{(t+1)}=\frac{\sum_{i=1}^n\hat \delta_i^{(t+1)}}{n}.
\label{eq:stepM}
\end{equation}
We follow an intuitive classification test criterion, classifying the points according to the mixture component where the volumes have the highest densities. We are mainly interested in identifying the feature points, and this proposed classification approach takes the estimated component with the highest intensity as the distribution of the feature point pattern. Additionally, for large $n$, the convergence of the $EM$ algorithm is good since it arrives at an approximately acceptable solution.

\subsection{Choosing the value of \textit{K}}\label{sec:choose}
\begin{figure}[h!]
\centering
\includegraphics[width=0.65\textwidth]{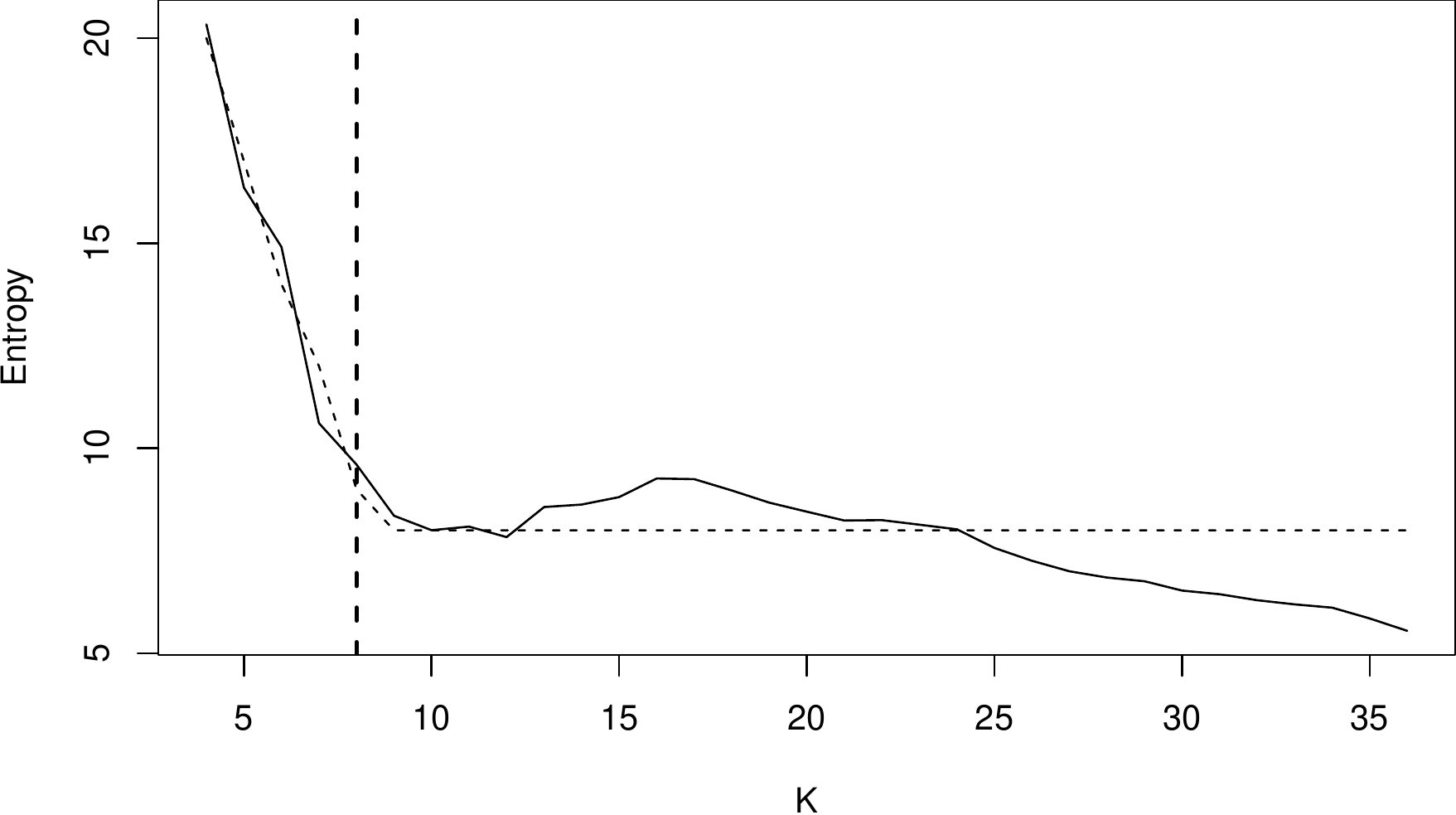}
\caption{\label{fig:seg} Estimated $K$ values for the simulated pattern on the \texttt{chicago} network in Figure \ref{fig:chicago}. The black line represents the observed entropies, and the dotted line represents the estimated segmented model. The vertical line indicates the estimated changepoint of $\hat K=8$.}
\end{figure}

The development mentioned above assumes a proper value of $K$ priorly chosen. The natural way to choose the suitable $K$-th neighbour is by analysing several increasing values of $K$ and then selecting that $K$, after which no improvement is found. However, in the literature, there are several methodological proposals for this target; in this work, we use an entropy-type measure of separation introduced in \cite{celeux1996entropy} given by  
\begin{equation}
    E = - \sum_{i=1}^{n} \delta_i \log_{2}(\delta_i),
    \label{eq:entropy}
\end{equation}
where $\delta_i$ are the probabilities of being in the first component of the mixture in equation \eqref{mix-dist}, that is, the feature. As stated by \cite{Byers1998}, plotting the entropies sequentially and looking for a levelling-off changepoint in the graph is an easy way to choose $K$. An example of this procedure is shown in Figure~\ref{fig:seg}, where the classification entropies for values of $K$ up to $35$ are plotted for the simulated point pattern on the \texttt{chicago} network in Figure~\ref{fig:chicago}.

The optimal $K$ can be automatically selected by fitting a segmented regression model as 
\begin{equation*}
\mathbb{E}\left[Y|x_i \right]=\beta+\gamma(x_i-{\psi})I(x_i<{\psi}),
\end{equation*}
where our interest is estimating a unique changepoint $\psi$,
after which the slope is constrained to be equal to zero. As depicted in Figure~\ref{fig:seg}, the observed response variable $Y$ is the entropy level, modelled as a function of the number of nearest neighbours $x$. We implemented this automatic option using the function \texttt{segmented} of the package \texttt{segmented} \citep{muggeo2008segmented}. We refer to \cite{muggeo2003estimating} for inferential details on segmented regression models. 

The following steps implement the classification procedure:
\begin{itemize}
\item[(1)] Choose a value of $K$ either by imputing a sought value or automatically applying a segmented regression model. Note that an upper bound for the $K$ domain should be fixed manually to a reasonable maximum number of neighbours.
\item[(2)] Find the $K$-th nearest neighbours distance and calculate the volume of the disc for each point in the point pattern using the shortest path distances computed on the given linear network.
\item[(3)] Apply the $EM$ algorithm for estimating $\lambda_1$, $\lambda_2$, and $p$.
\item[(4)] Classify the points as features according to whether they have a higher density under the mixture's components.
\end{itemize}

\section{Simulation study}\label{sec:simulations}
This section aims to study the proposed method's performance in terms of classification rates, considering different scenarios and feature shapes. We generate clutter in the whole network in all scenarios, and the feature is simulated and constrained to a sub-network. To explore the classification procedure for different networks, we use both the \texttt{chicago} and the \texttt{dendrite} networks \citep{spatstat.data} in the simulation study, which represents linear networks with and without cycles. In the third scenario, we use the \texttt{chicago} network, taking two nested sub-networks and generating clutter with two different intensities. The fourth network is the Antonio Narino network \citep{Moncada2018}, which has a homogeneous structure, and we simulate features in two disjoint sub-networks corresponding  to two separate roads.

We show the results of the classification method in terms of true-positive rate (TPR), false-positive rate (FPR), and accuracy (ACC), averaging over 100 simulated point patterns generated with $\mathbb{E}[n_c]$ and $\mathbb{E}[n_f]$ expected number of points for clutter and feature, respectively. In addition, the $\lambda_c$ and $\lambda_f$ intensities are reported for clutter and feature. The accuracy is the proportion of correct predictions (both true positives and negatives) among the total number of cases examined. The tables also report the average of the 100 estimates of $K$ by the segmented regression model for the simulated point patterns in each design ($\bar K$) and its standard deviation (sd). We further compare the detection with $K$ equal to $\{5, 10\}$ and the automatically selected by the segmented regression model considered in Section \ref{sec:choose}. The rates are computed accordingly. 

\subsection{Estimating feature in presence of clutter and cycles}\label{subsec:chicago}
\begin{figure}[h!]
\centering
\subfloat[Simulate pattern.]
{\includegraphics[width=0.4\textwidth]{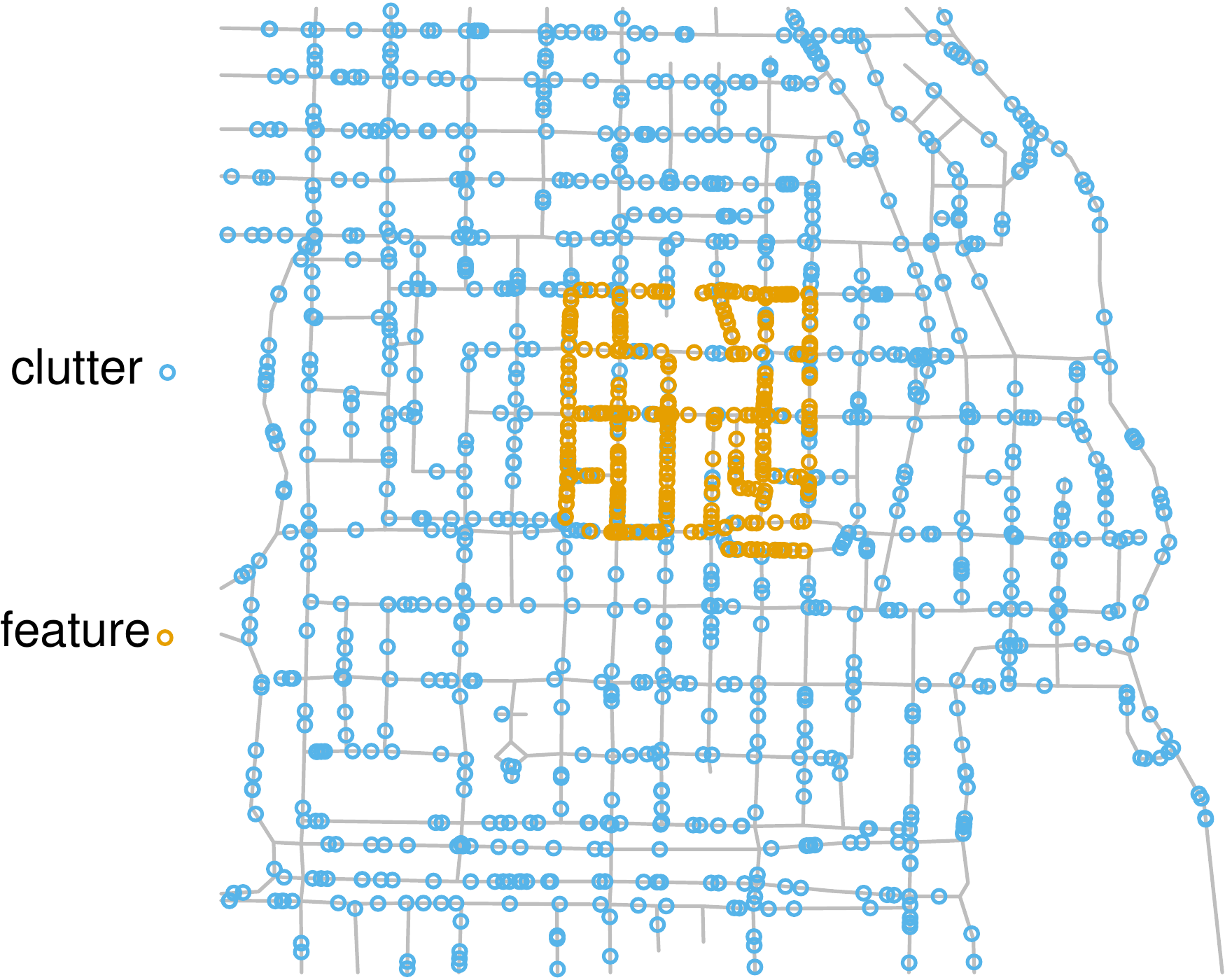}\label{fig:simulatechicago}}\qquad\qquad
\subfloat[Classified pattern.]{\includegraphics[width=0.4\textwidth]{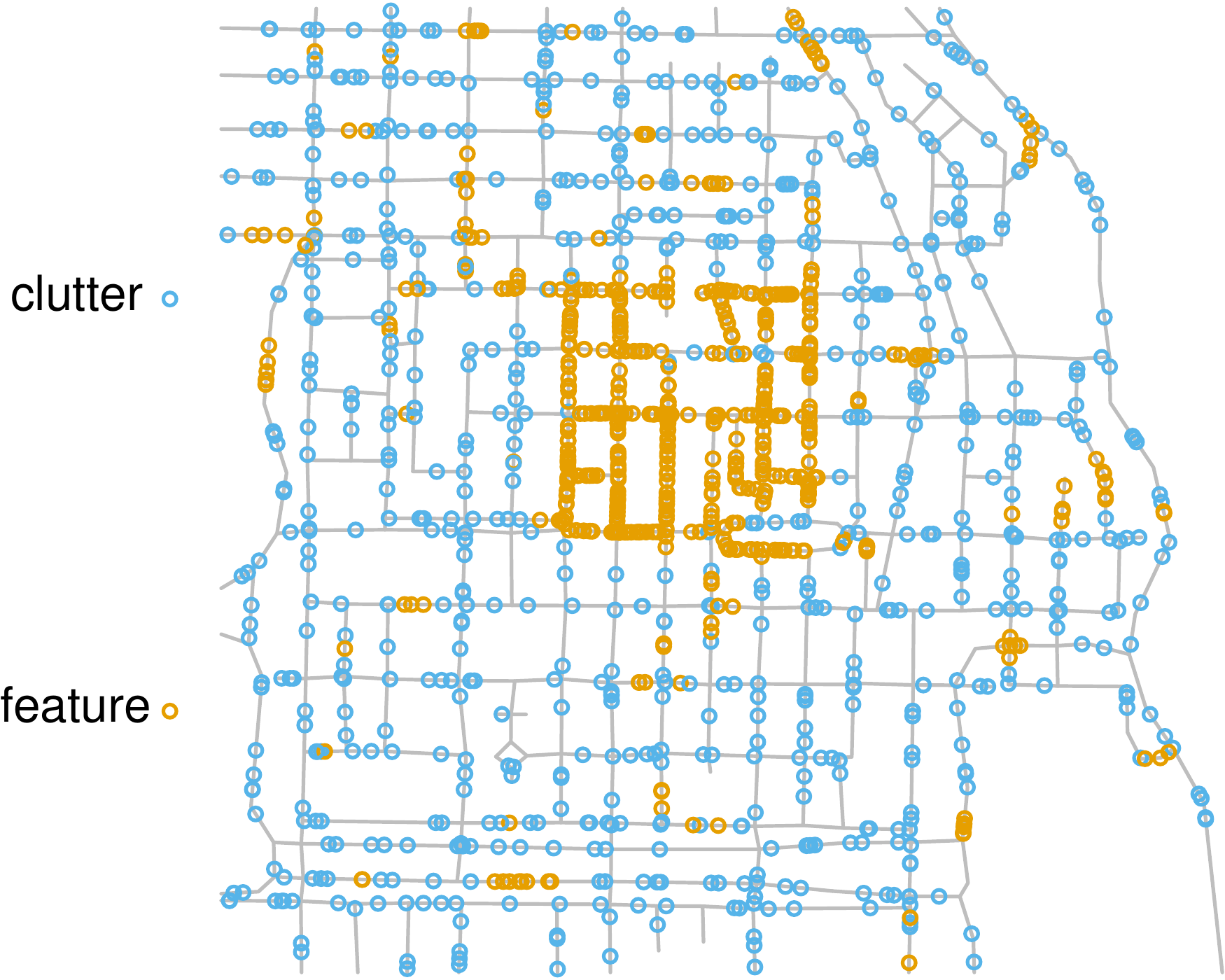}\label{fig:classifchicago}}
\caption{\textit{Left:} Simulated point pattern on the \texttt{chicago} linear network. The clutter and feature are simulated from a Poisson point process homogeneous with $\lambda_{c}=0.032$, $\mathbb{E}[n_{c}]=1000$ and $\lambda_{f}=0.100$ and $\mathbb{E}[n_f]=300$ respectively. \textit{Right:} Randomly selected classified point pattern using the proposed classification method.}\label{fig:ChicagoSimClas}
\end{figure}

The linear network \texttt{chicago} described in Section \ref{sec:approach} is a network with loops given with edges connecting vertexes to themselves. As mentioned in that section, we generate point patterns of feature and clutter as displayed in Figure~\ref{fig:simulatechicago}, and report the outcomes in Table \ref{tab:chicago}.

From Table \ref{tab:chicago}, for designs 1 and 2, where $\lambda_c < \lambda_f$, it is possible to notice that the classification method performs well in terms of true and false positive rates for $K=5$, $K=10$ and $\hat K$. For designs 3 and 4, where $\lambda_c \approx \lambda_f$, the classification method performs well in true-positive rates for $K=5$, $K=10$ and $\hat K$, but there is a significant increase in false-positive rates. In designs 5 and 6, where $\lambda_c > \lambda_f$, the classification method has lower performance for $K=5$, $K=10$ and $\hat K$, with a decrease in the true-positive rates and an increase in the false-positive rates. The True Positive rates always being higher than false-positive rates. In general, the accuracy shows that the best performance is achieved when the expected number of feature points approaches the expected number of clutter points, that is when their ratio comes close to 1. The worst performance is when this ratio is close to 0. In the former case, there is little clutter compared to the number of features, and the classification method quickly detects the features as in design 1; whereas in the latter case, the amount of clutter is too large, and the features are more hidden to be detected, as in design 6. The estimated $\hat{K}$ selected by minimizing the entropy level defined in equation \eqref{eq:entropy} has similar classification rates as with $K=5$ and $K=10$ for every design showing that values of $K$ within a reasonable range, tend to perform similarly.
\begin{table}[H]
\centering
\scalebox{0.7}{
\begin{tabular}{cccccccc|ccc}
 \toprule
&&&&&&&&&$K$&\\\cmidrule(l){9-11}
Design&$\lambda_c$&$\lambda_f$&$\mathbb{E}[n_c]$& $\mathbb{E}[n_f]$&$\bar{K}$&sd(K)&Rate& 5 & 10 & $\hat{K}$\\ 
  \midrule
1&0.032&0.100& 1000& 300& 4.19& 0.42& TPR& 0.962 & 0.952 & 0.963 \\ 
&&& & & & & FPR& 0.241 & 0.134 & 0.274 \\ 
&&& & & & &ACC&  0.805 & 0.886 & 0.780 \\
   \midrule
2&0.032&0.067& 1000& 200&  4.81& 0.49& TPR& 0.944 & 0.931 & 0.946 \\ 
&&& &  & & &FPR& 0.356 & 0.189 & 0.366 \\ 
&&& &  & & &ACC&   0.694 & 0.831 & 0.686 \\ 
   \midrule
3&0.032&0.033& 1000 &100 & 7.34& 1.62& TPR& 0.896 & 0.910 & 0.900 \\ 
&&&   &  & & & FPR&  0.490 & 0.398 & 0.436 \\ 
 &&&  &   & & &ACC&   0.545 & 0.629 & 0.594 \\ 
   \midrule
4&0.016&0.017& 500 &50 & 7.26& 1.15& TPR& 0.904 & 0.891 & 0.905 \\ 
&&&   &  & & & FPR& 0.502 & 0.379 & 0.448 \\ 
 &&&  &  & & &ACC&   0.535 & 0.646 & 0.585 \\ 
     \midrule
5&0.064&0.017& 2000 &50 & 4.42& 0.61& TPR& 0.715 & 0.733 & 0.721 \\ 
&&&   & & & & FPR& 0.543 & 0.512 & 0.555 \\ 
 &&&  & &  & &ACC&   0.464 & 0.494 & 0.452 \\
\midrule
6&0.128&0.017& 4000 &50 & 5.42& 0.50& TPR& 0.683 & 0.642 & 0.680 \\ 
&&&   & & & & FPR& 0.578 & 0.524 & 0.574 \\ 
 &&&  &  & & &ACC& 0.425 & 0.478 & 0.429 \\
  \bottomrule
\end{tabular}}
\caption{\label{tab:chicago} Classification rates averaged over 100 simulated point patterns generated on the \texttt{chicago} linear network with $\lambda_c$ and $\lambda_f$ intensities and $\mathbb{E}[n_c]$ and $\mathbb{E}[n_f]$ expected number of points for clutter and feature.}
\end{table}

\subsection{Estimating feature in presence of clutter and without cycles}\label{subsec:dendrite}
\begin{figure}[ht!]
\centering
\subfloat[Simulate pattern.]
{\includegraphics[width=0.45\textwidth]{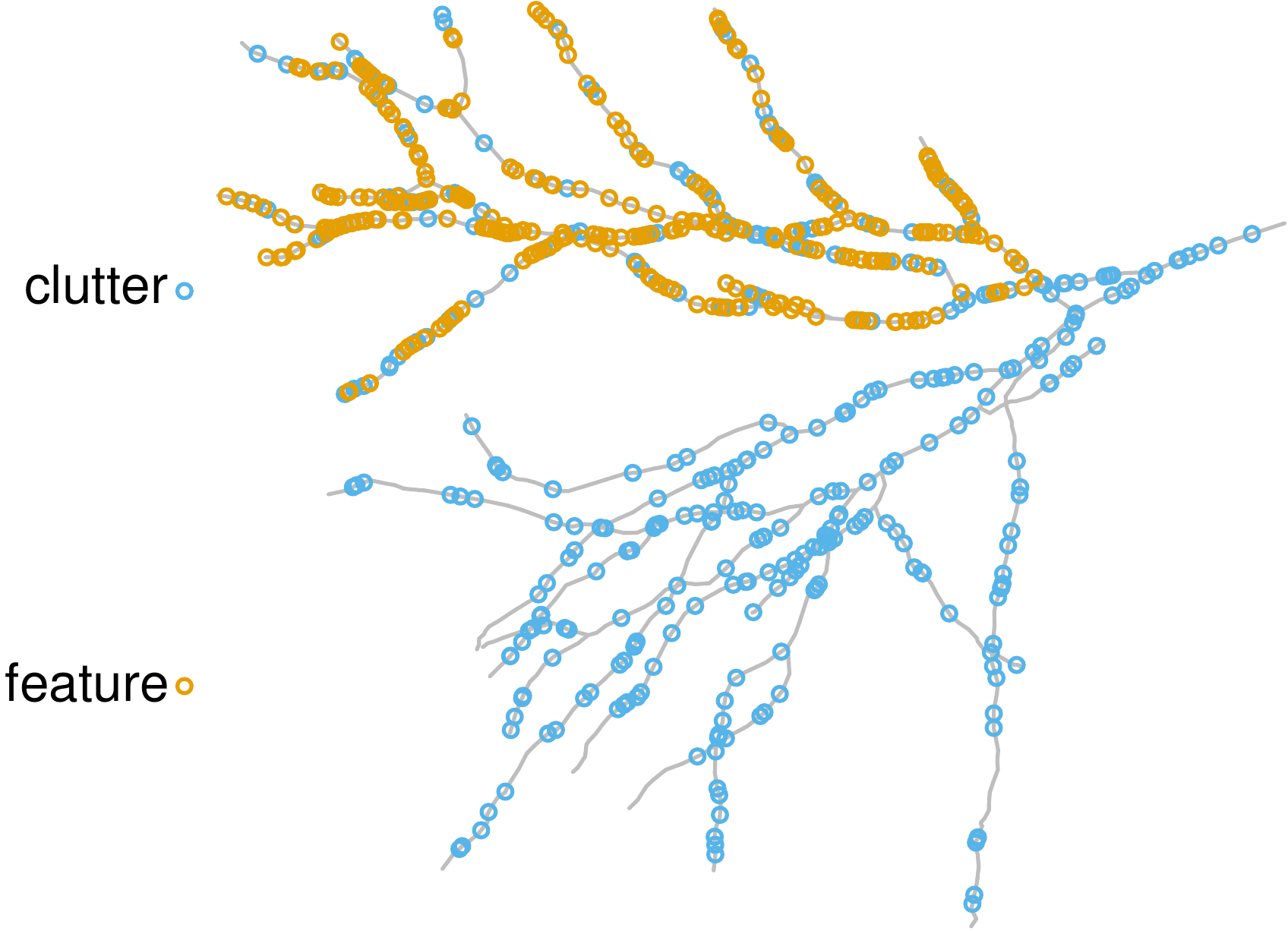}\label{fig:Dendrite}}\qquad\qquad
\subfloat[Classified pattern.]{\includegraphics[width=0.45\textwidth]{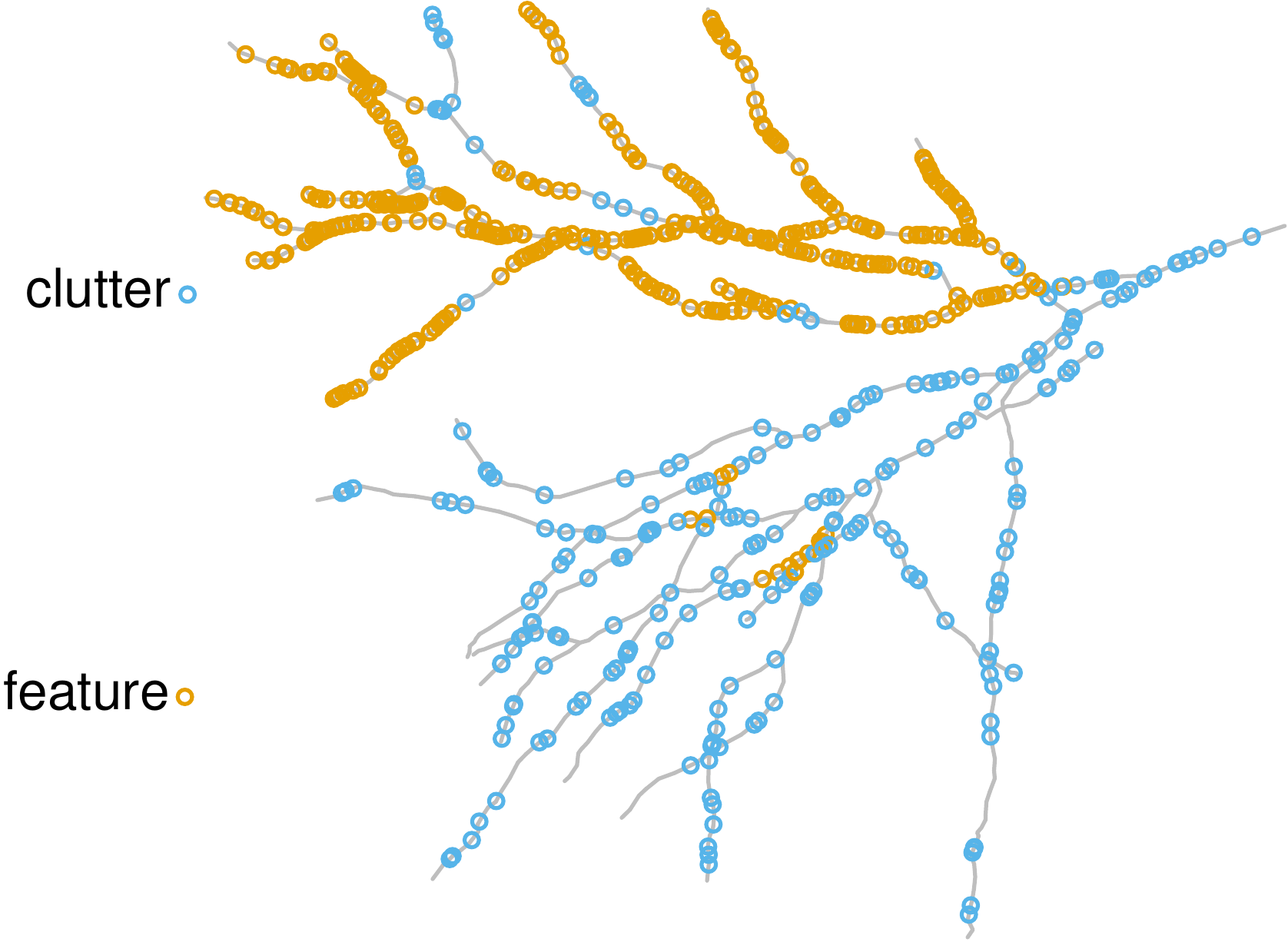}\label{fig:classifDendrite}}
\caption{\textit{Left:} Simulated point pattern on the \texttt{dendrite} linear network. The clutter and feature are simulated from a Poisson point process homogeneous with $\lambda_{c}=0.207$, $\mathbb{E}[n_c]=400$ and $\lambda_{f}=0.386$, $\mathbb{E}[n_f]=300$ respectively. \textit{Right:} Randomly selected classified point pattern using the proposed classification method.}\label{fig:dendrite}
\end{figure}

The linear network \texttt{dendrite} is the dendrite network of a neuron (nerve cell) recorded by the Kosik Lab (UCSB) \citep{jbmk2013} and available in the package \texttt{spatstat.data} \citep{spatstat.data}. The \texttt{dendrite} network has 640 intersections, 639 uninterrupted segments and a total length of 1934 $\mu$m (1.93 mm) \citep{BJN2014}, while the sub-network we choose has 245 intersections, 243 uninterrupted segments and a total length of 778 $\mu$m (0.78 mm). We generate feature and clutter points on the network branches, as shown in Figure~\ref{fig:Dendrite}.

\begin{table}[ht!]
\centering
\scalebox{0.7}{
\begin{tabular}{cccccccc|ccc}
 \toprule
&&&&&&&&&$K$&\\\cmidrule(l){9-11}
Design&$\lambda_c$&$\lambda_f$&$\mathbb{E}[n_c]$& $\mathbb{E}[n_f]$&$\bar{K}$&sd(K)&Rate& 5 & 10 & $\hat{K}$ \\ 
  \midrule
1&0.207&0.386& 400& 300& 7.98& 0.67& TPR& 0.910 & 0.956 &0.941  \\ 
&&&&& & & FPR& 0.478 & 0.428 & 0.438 \\ 
&&&&& & &ACC&  0.688 & 0.737 & 0.725 \\ 
   \midrule
2&0.207&0.257& 400& 200 & 8.78& 0.75& TPR& 0.863 & 0.905 & 0.894 \\ 
&&&&& &  &FPR& 0.504 & 0.448 & 0.453 \\ 
&&&&& &  &ACC&   0.619 & 0.670 & 0.664 \\ 
   \midrule
3&0.388&0.386& 750 &300 & 12.64& 1.09& TPR& 0.842 & 0.891 & 0.910 \\ 
&&&&&   &  & FPR& 0.532 & 0.473 & 0.453 \\ 
 &&&&&  &   &ACC&   0.575 & 0.631 & 0.650 \\ 
   \midrule
4&0.259&0.257& 500  &200 & 10.30& 1.21& TPR& 0.849 & 0.892 & 0.892 \\ 
&&&&&   &  & FPR& 0.539 & 0.476 & 0.474 \\ 
 &&&&&  &   &ACC&  0.571 & 0.628 & 0.630 \\
 \midrule
5&0.259&0.193& 500 &150 & 11.02& 1.47& TPR& 0.802 & 0.840 & 0.846 \\ 
&&&&&   &  & FPR&  0.541 & 0.485 & 0.474 \\ 
 &&&&&  &   &ACC&   0.538 & 0.590 & 0.600 \\
     \midrule
6&0.388&0.032& 750  &25 & 11.91& 1.91& TPR& 0.569 & 0.559 & 0.548 \\ 
&&&&&   &  & FPR&  0.560 & 0.520 & 0.505 \\ 
 &&&&&  &   &ACC&  0.444 & 0.483 & 0.497 \\
  \bottomrule
\end{tabular}}
\caption{\label{tab:dendrite} Classification rates averaged over 100 simulated point patterns generated on the \texttt{dendrite} linear network with $\lambda_c$ and $\lambda_f$ intensities and $\mathbb{E}[n_c]$ and $\mathbb{E}[n_f]$ expected number of points for clutter and feature.}
\end{table}

 We report the results of the classification procedure in Table \ref{tab:dendrite}. For designs 1 and 2, where $\lambda_c < \lambda_f$, it is possible to notice that the classification method performs well in terms of true-positive rates for $K=5$, $K=10$ and $\hat K$, but has high false-positive rates. For designs 3 and 4, where $\lambda_c \approx \lambda_f$, and design 5, where $\lambda_c > \lambda_f$, the classification method also performs well in terms of true-positive rates for $K=5$, $K=10$ and $\hat K$, and has an increase in false-positive rates. For design 6, where $\lambda_c > \lambda_f$, the classification method has low performance in terms of true and false positive rates for $K=5$, $K=10$ and $\hat K$. The true positive rates always are higher than false-positive rates. The method performs better when the ratio between the expected number of features and the expected number of clutter approaches one. This outcome lines up with the previous experiments for networks with loops. 

\subsection{Estimating feature in presence of two nested clutters}\label{subsec:chicago2}
We consider an even more complex scenario where the features could be somehow hidden by different intensities found in the network. Therefore, to show the procedure's usefulness, we simulate from a more complex setting dealing with the mixture of multiple intensities of a homogeneous Poisson process on the observed point pattern. We use again the \texttt{chicago} linear network and the feature sub-network defined in Section \ref{subsec:chicago}.
\begin{figure}[h!]
\centering
\subfloat[Simulate pattern.]
{\includegraphics[width=0.4\textwidth]{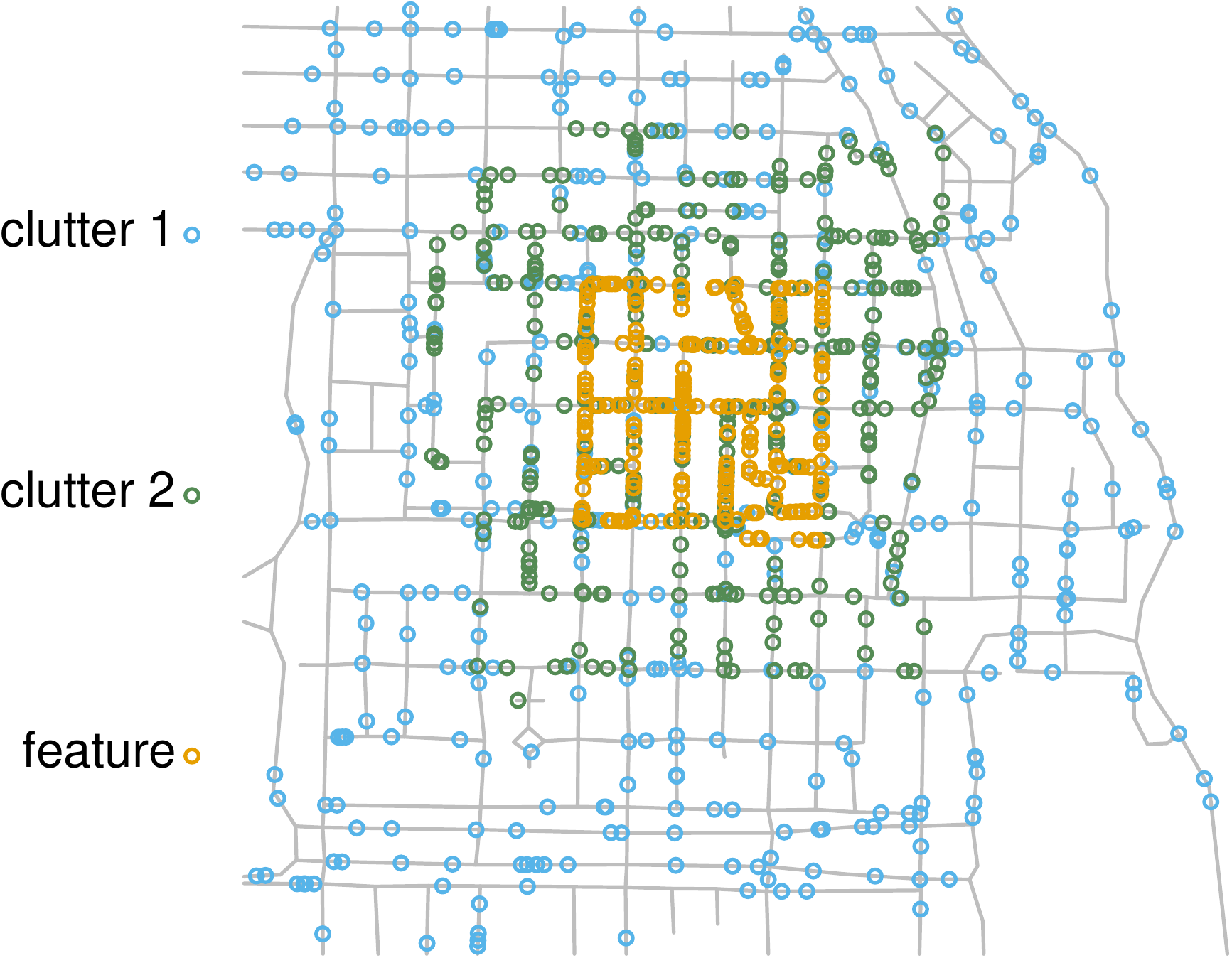}\label{fig:simulatechicago2}}\qquad\qquad
\subfloat[Classified pattern.]{\includegraphics[width=0.4\textwidth]{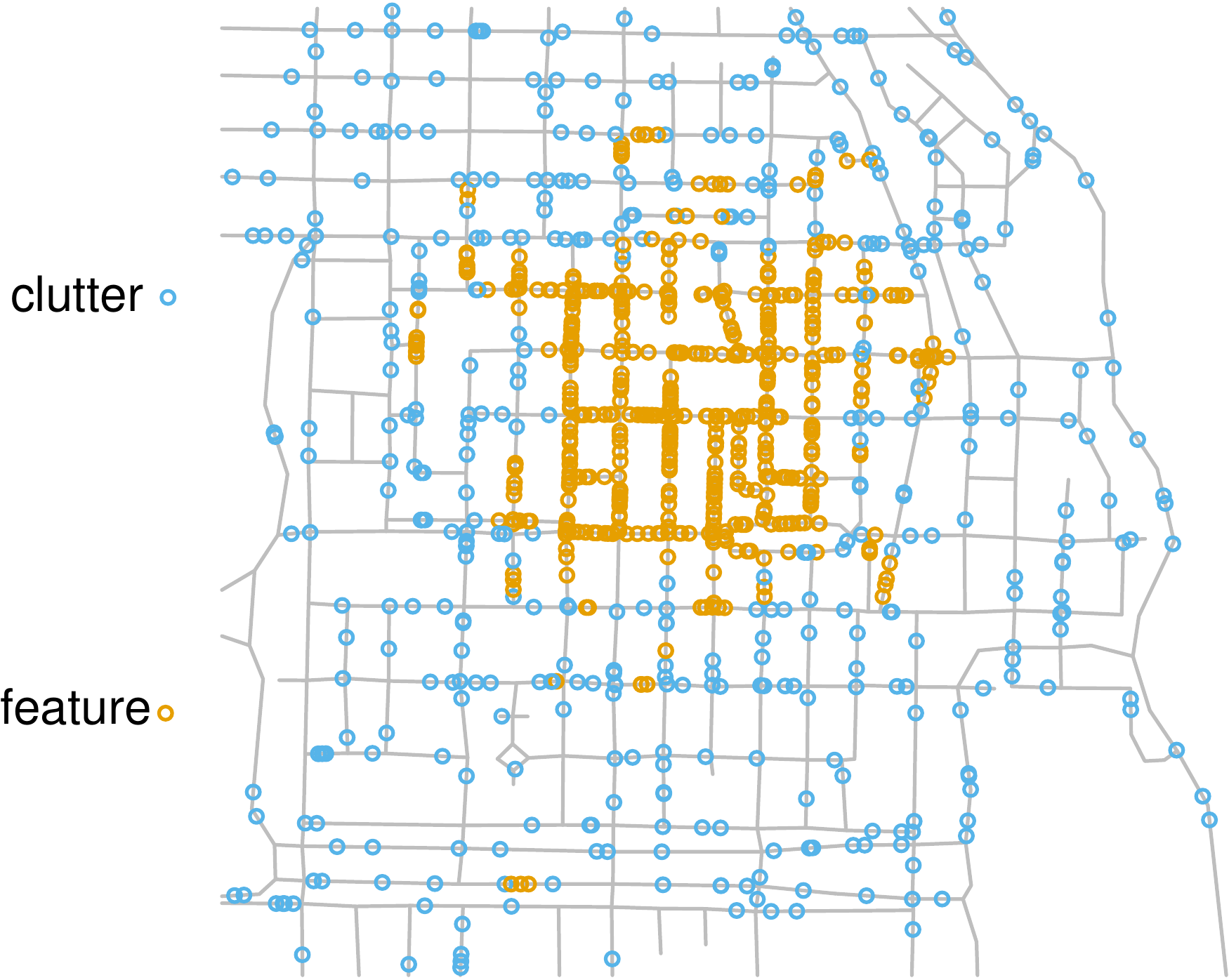}\label{fig:classifchicago2}}
\caption{\textit{Left:} Simulated point pattern on the \texttt{chicago} linear network with two nested sub-networks. First and second clutter are simulated from a homogeneous Poisson point process with $\lambda_{c_1}=0.013$, $\mathbb{E}[n_{c_1}]=400$ in the whole \texttt{chicago} network and  $\lambda_{c_2}=0.026$, $\mathbb{E}[n_{c_2}]=300$ in the nested sub-network containing the feature. The feature is simulated from a homogeneous Poisson point process with $\lambda_{f}=0.067$ and $\mathbb{E}[n_f]=200$. \textit{Right:} Randomly selected classified point pattern using the proposed classification method.}\label{fig:multi}
\end{figure}

In this case, we consider a third intensity value in a larger nested sub-network than the feature sub-networks as shown in Figure~\ref{fig:simulatechicago2}; the largest nested sub-network of the \texttt{chicago} has 130 intersections, 199 uninterrupted segments and a total length of 11731 feet. Hence, we simulate two distinct Poisson point patterns with constant intensities in the two nested sub-networks and superimpose them to a point pattern generated on the whole network. In other words, we  create a second clutter in the larger sub-network and the feature points in the smaller one.
\begin{table}[ht!]
\centering 
\scalebox{0.65}{
\begin{tabular}{cccccccccc|ccc}
 \toprule
&&&&&&&&&&&$K$&\\\cmidrule(l){11-13} 
Design&$\lambda_{c_1}$&$\lambda_{c_2}$&$\lambda_f$&$\mathbb{E}[n_{c_1}]$& $\mathbb{E}[n_{c_2}]$&$\mathbb{E}[n_f]$&$\bar{K}$&sd(K)&Rate& 5 & 10 & $\hat{K}$\\   
\midrule
1&0.013&0.026&0.067& 400& 300&200& 4.19& 0.42& TPR&  0.988 & 0.998 & 0.980 \\ 
&&&&&& & && FPR& 0.423 & 0.446 & 0.426 \\ 
&&&&&& && &ACC&  0.667 & 0.651 & 0.663 \\
\midrule
2&0.032&0.051&0.067& 1000&600  &200 & 4.89& 0.31& TPR&  0.972 & 0.994 & 0.971  \\ 
&&&&&& & && FPR& 0.503 & 0.509 & 0.505 \\ 
&&&&&& && &ACC& 0.551 & 0.547 & 0.549  \\
     \midrule
3&0.032&0.085&0.067& 1000 &1000& 200 & 4.56& 0.50& TPR&  0.984 & 0.998 & 0.981 \\ 
&&&&&& & && FPR& 0.622 & 0.646 & 0.624 \\ 
&&&&&& && &ACC&  0.433 & 0.413 & 0.432 \\  
  \midrule
4&0.032&0.085&0.017& 1000& 1000&50& 4.37& 0.49& TPR&  0.942 & 0.987 & 0.935 \\ 
&&&&&& & && FPR& 0.645 & 0.673 & 0.647 \\ 
&&&&&& && &ACC&   0.369 & 0.343 & 0.368 \\ 
\midrule
5&0.032&0.034&0.033& 1000 &400& 100 & 4.87& 0.34& TPR&  0.943 & 0.985 & 0.943 \\ 
 &&&&&& & && FPR& 0.531 & 0.523 & 0.533 \\ 
&&&&&& && &ACC&  0.500 & 0.510 & 0.498 \\ 
     \midrule
6&0.064&0.068&0.067& 2000 &800& 200 & 5.13& 0.34& TPR& 0.952 &
0.977 & 0.953 \\ 
&&&&&& & && FPR& 0.535 & 0.487 & 0.532 \\ 
&&&&&& && &ACC&  0.499 & 0.544 & 0.502 \\
   \midrule
7&0.064&0.034&0.134& 2000&400 &400 & 5.03& 0.17& TPR& 0.975 & 0.974 & 0.976 \\ 
&&&&&& & && FPR& 0.402 & 0.281 & 0.401 \\ 
&&&&&& && &ACC&  0.653 & 0.756 & 0.654 \\ 
   \midrule
8&0.128&0.026&0.033& 4000&300  &100  & 5.84& 0.42& TPR& 0.810 & 0.816 & 0.815 \\ 
&&&&&& & && FPR& 0.587 & 0.526 & 0.576 \\ 
&&&&&& && &ACC& 0.422 & 0.482 & 0.433 \\
     \midrule
9&0.064&0.026&0.017& 2000 &300& 50 & 4.83& 0.47& TPR& 0.820 & 0.847 & 0.823 \\ 
&&&&&& & && FPR& 0.561 & 0.517 & 0.567 \\ 
&&&&&& && &ACC&  0.447 & 0.491 & 0.442 \\
 \midrule
10&0.128&0.051&0.017& 4000 &600& 50 & 5.97& 0.33& TPR& 0.811 & 0.815 & 0.816 \\ 
 &&&&&& & && FPR& 0.604 & 0.544 & 0.592 \\ 
&&&&&& && &ACC&  0.401 & 0.460 & 0.412 \\ 
 \bottomrule
\end{tabular}}
\caption{\label{tab:chicago_multi} Classification rates averaged over 100 simulated point patterns generated on the \texttt{chicago} linear network with two nested sub-networks with $\lambda_{c_1}$, $\lambda_{c_2}$ and $\lambda_f$ intensities and $\mathbb{E}[n_{c_1}]$, $\mathbb{E}[n_{c_2}]$ and $\mathbb{E}[n_f]$ expected numbers of points for clutter one, clutter two and feature, respectively.}
\end{table}

Table \ref{tab:chicago_multi} shows the results of the feature classification procedure for the \texttt{chicago} network with two nested patterns of clutter. As in the previous settings, the results are computed for different combinations of the intensities $\lambda_{c_1}$, $\lambda_{c_2}$ and $\lambda_f$. In designs 1 and 2, we take $\lambda_{c_1} < \lambda_{c_2} < \lambda_f$. In designs 3 and 4, we take $\lambda_{c_1} < \lambda_{c_2}>\lambda_f$ for $\lambda_{c_1} < \lambda_f$ in design 3 and $\lambda_{c_1} > \lambda_f$ for design 4. In designs 5 and 6, we take $\lambda_{c_1} \approx \lambda_{c_2} \approx \lambda_f$. In designs 7 and 8, we take $\lambda_{c_1} > \lambda_{c_2}<\lambda_f$ for $\lambda_{c_1} < \lambda_f$ in design 7 and with $\lambda_{c_1} > \lambda_f$ in design 8. In designs 9 and 10, we take $\lambda_{c_1} > \lambda_{c_2} > \lambda_f$.
Looking at the classification rate in Table \ref{tab:chicago_multi}, we can see a very good method performance for designs 1 to 7 in terms of true positive rates for $\hat K$, $K=5$ and $K=10$. In designs 8, 9 and 10 decrease the true positive rates, although it is still good. In all designs, we can see different variations of false-positive rates for $\hat K$, $K=5$ and $K=10$, but consistently lower than true-positive rates. In this scenario, we can also notice that the method performance is better when the ratio between the expected number of features and the expected number of clutter approaches one.

\subsection{Estimating two disjoint features in presence of clutter}\label{subsec:antonio}
To highlight the flexibility of the classification method in terms of feature shape, we simulate a down-to-earth scenario on the streets of Antonio Narino locality in Bogota, Colombia \citep{Moncada2018}. 
\begin{figure}[H]
\centering
\subfloat[Simulate pattern.]
{\includegraphics[width=0.6\textwidth]{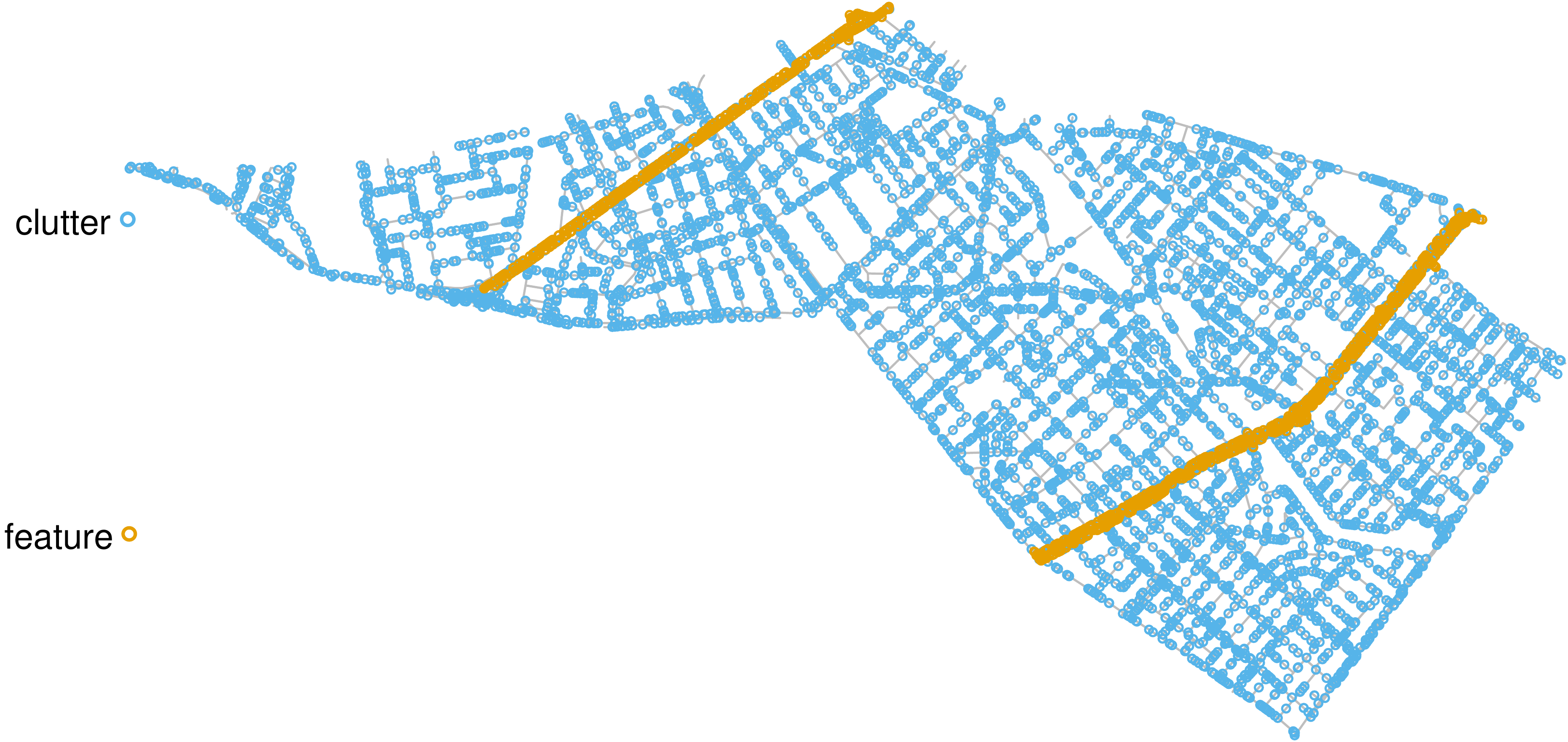}\label{fig:antonio}}\qquad\qquad
\subfloat[Classified pattern.]{\includegraphics[width=0.6\textwidth]{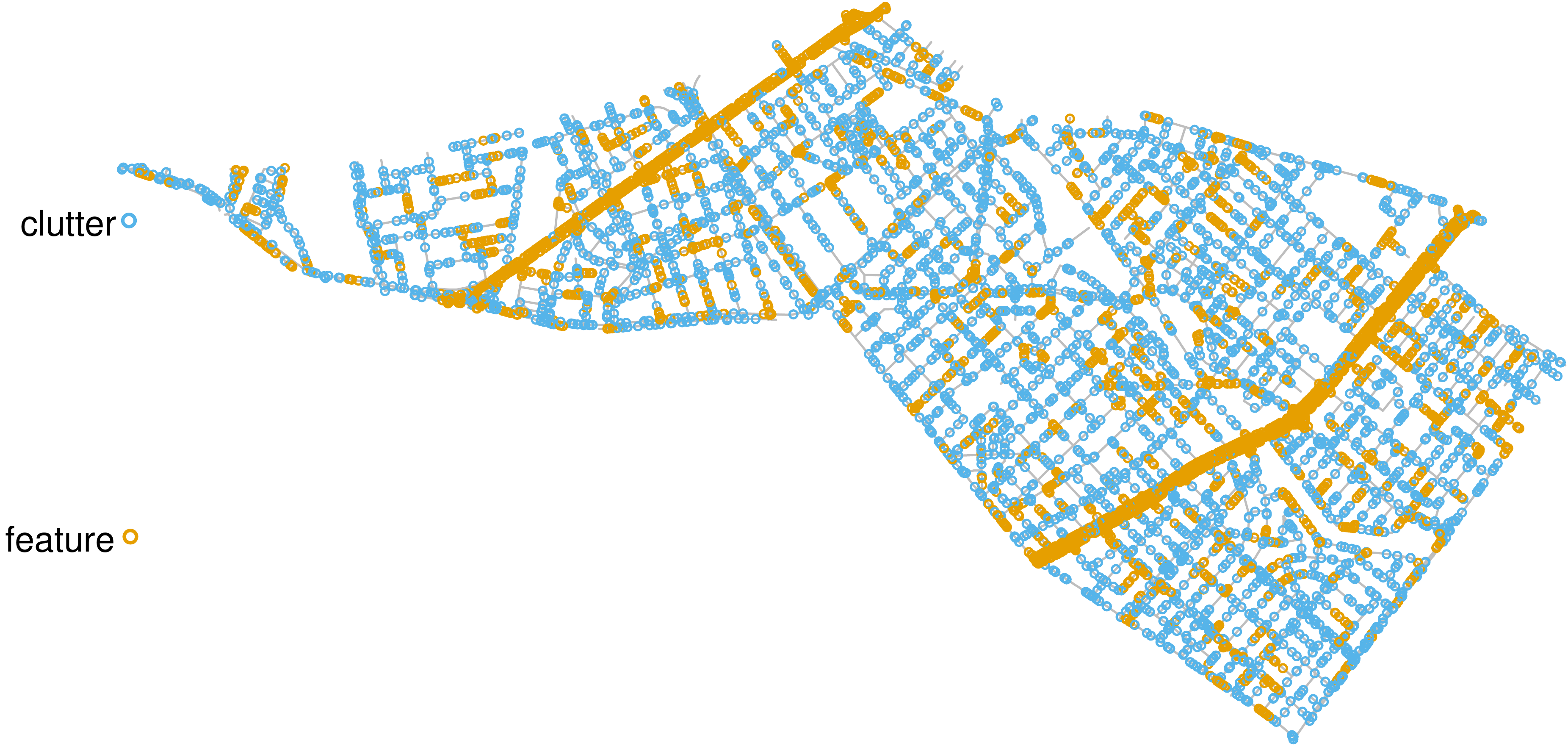}\label{fig:classifantonio}}
\caption{\textit{Left:} Simulated point pattern on the Antonio Nariño linear network. The clutter is simulated from a homogeneous Poisson point process with $\lambda_{c}=0.047$ and $\mathbb{E}[n_c] = 6000$. Both features are simulated from homogeneous Poisson point processes on two disjoint streets with $\lambda_{f_1}=0.072$, $\mathbb{E}[n_{f_1}] = 600$ and $\lambda_{f_2}=0.109$, $\mathbb{E}[n_{f_2}] = 400$. \textit{Right:} Randomly selected classified point pattern using the proposed classification method.}\label{fig:AntonioSimClas}
\end{figure}

Antonio Narino is one of the eighteen zones, called Localidades, into which the metropolitan area of Bogota is divided. The road network of Antonio Narino locality has 2062 intersections, 2706 uninterrupted segments and a total length of 128.69 km. The feature points are simulated on two of the longest roads, as shown in Figure~\ref{fig:antonio}. The first road (right) is a sub-network with 237 intersections, 254 uninterrupted segments and a total length of 8.32 km and the second road (left) is a sub-network with 106 intersections, 112 uninterrupted segments and a total length of 3.68 km.
\begin{table}[ht!]
\centering 
\scalebox{0.7}{
\begin{tabular}{cccccccccc|ccc}
\toprule
&&&&&&&&&&&$K$&\\\cmidrule(l){11-13} 
Design&$\lambda_c$&$\lambda_{f_1}$&$\lambda_{f_2}$&$\mathbb{E}[n_c]$& $\mathbb{E}[n_{f_1}]$&$\mathbb{E}[n_{f_2}]$&$\bar{K}$&sd(K)&Rate& 5 & 10 & $\hat{K}$\\ 
 \midrule
1&0.047&0.072&0.109& 6000&600  &400  & 5.07& 0.26& TPR& 0.966 & 0.904 & 0.967 \\ 
&&&&&& & && FPR& 0.432 & 0.189 & 0.430 \\ 
&&&&&& && &ACC& 0.625 & 0.825 & 0.627 \\  
\midrule
2&0.023&0.036&0.054& 3000& 300&200& 5.58& 0.65& TPR& 0.969 & 0.924 & 0.962 \\ 
&&&&&& & && FPR& 0.445 & 0.224 & 0.418 \\ 
&&&&&& && &ACC&  0.615 & 0.798 & 0.637 \\ 
     \midrule
3&0.047&0.072&0.054& 6000&600  &200 & 5.84& 0.56& TPR&  0.956 & 0.903 & 0.954 \\ 
&&&&&& & && FPR& 0.481 & 0.259 & 0.450 \\ 
&&&&&& && &ACC& 0.570 & 0.760 & 0.597 \\  
  \midrule
4&0.047&0.072&0.014& 6000&600  &50 & 6.63& 0.76& TPR&  0.949 & 0.886 & 0.948 \\ 
&&&&&& & && FPR& 0.492 & 0.239 & 0.447 \\ 
&&&&&& && &ACC& 0.551 & 0.773 & 0.592 \\ 
\midrule
5&0.093&0.093&0.094& 12000&775  &345 & 6.32& 0.63& TPR& 0.935 & 0.928 & 0.942 \\ 
&&&&&& & && FPR& 0.539 & 0.395 & 0.510 \\ 
&&&&&& && &ACC& 0.501 & 0.633 & 0.529 \\
   \midrule
6&0.035&0.036&0.034& 4500&300 &125 & 7.29& 0.89& TPR& 0.922 & 0.940 & 0.937 \\ 
&&&&&& & && FPR& 0.513 & 0.427 & 0.479 \\ 
&&&&&& && &ACC&  0.525 & 0.604 & 0.557 \\ 
     \midrule
7&0.047&0.024 & 0.094& 6000 &200& 345 & 8.02& 0.64& TPR&  0.913 & 0.776 & 0.853  \\ 
 &&&&&& & && FPR& 0.503 & 0.218 & 0.356 \\ 
&&&&&& && &ACC&  0.531 & 0.781 & 0.661 \\ 
 \midrule
8&0.047&0.024 & 0.034& 6000 &200& 125  & 5.18& 0.73& TPR&  0.839 & 0.880 & 0.841 \\ 
 &&&&&& & && FPR& 0.540 & 0.497 & 0.538 \\ 
&&&&&& && &ACC& 0.479 & 0.522 & 0.481  \\
     \midrule
9&0.047&0.024 & 0.014& 6000 &200& 50 & 4.92& 0.61& TPR&  0.807 & 0.855 & 0.806  \\ 
 &&&&&& & && FPR& 0.546 & 0.514 & 0.547 \\ 
&&&&&& && &ACC&  0.468 & 0.501 & 0.467 \\ 
     \midrule
10&0.093&0.024 & 0.014& 12000 &200& 50 & 4.98& 0.14& TPR& 0.713 & 0.745 & 0.713  \\ 
&&&&&& & && FPR& 0.572 & 0.527 & 0.573  \\ 
&&&&&& && &ACC& 0.434 & 0.478 & 0.433 \\
  \bottomrule
\end{tabular}}
\caption{\label{tab:antonio} Classification rates averaged over 100 simulated point patterns generated on the Antonio Nariño linear network with $\lambda_c$, $\lambda_{f_1}$ and $\lambda_{f_2}$ intensities and $\mathbb{E}[n_c]$, $\mathbb{E}[n_{f_1}]$ and $\mathbb{E}[n_{f_2}]$ expected number of points for clutter, feature one and feature two, respectively.}
\end{table}

The report of the results of the classification procedure is shown in Table \ref{tab:antonio}. In the first road, we simulate a pattern of feature (feature 1) with expected number $\mathbb{E}[n_{f_1}]$ and intensity $\lambda_{f_1}$. Likewise, in the second road, we simulate other pattern of feature (feature 2) with expected number $\mathbb{E}[n_{f_2}]$ and intensity $\lambda_{f_2}$. These results are shown for different combinations of intensities $\lambda_c$, $\lambda_{f_1}$ and $\lambda_{f_2}$. In designs 1 and 2, we take $\lambda_c < \lambda_{f_1}<\lambda_{f_2}$. In designs 3 and 4, we take $\lambda_c < \lambda_{f_1}>\lambda_{f_2}$ with $\lambda_c < \lambda_{f_2}$ in design 3 and $\lambda_c > \lambda_{f_2}$ in design 4. In designs 5 and 6, we take $\lambda_c \approx \lambda_{f_1} \approx \lambda_{f_2}$. In designs 7 and 8, we take $\lambda_c > \lambda_{f_1}<\lambda_{f_2}$ with $\lambda_c < \lambda_{f_2}$ in design 7 and $\lambda_c > \lambda_{f_2}$ in design 8. In designs 9 and 10, we take $\lambda_c > \lambda_{f_1} > \lambda_{f_2}$.
The classification rates of the procedure in Table \ref{tab:antonio} 
have a similar interpretation to those in the previous scenario in Section \ref{subsec:chicago2}, showing a good classification method performance in all designs from 1 to 10.

\section{Identifying high- and low-density road traffic accident in two cities of Colombia}\label{sec:bogtraffic}
A traffic accident is an involuntary event generated by at least one vehicle in motion on roads in which there is damage to vehicles and objects and injuries or death to the people involved \citep{world2018global}. According to \cite{world2018global}, traffic accidents are the eighth cause of death worldwide and the leading cause of death for people between 5 and 29 years of age. Particularly in Colombia, \cite{itfRoadSafety} reports that between 2010 and 2018, the trend for the number of dead people in road accidents has been upward, and the annual number of these deaths increased by nearly 30\%. Based on the report of the National Road Safety Agency of Colombia \cite{ansvcol}, in the period 2016 to 2019, the number of dead people in road accidents annually remained between 3.4\% and 3.7\% of the total traffic accidents. When a traffic accident occurs, a chain of actions of different private and public entities begins to clarify what happened and determine the cause of this traffic accident. This means that the investigation of the reasons for which the accident is generated begins after its occurrence. Some factors that generally produce a traffic accident are human, environmental or mechanical. These factors change according to the country or city where the accidents occur; therefore, traffic accidents can be seen as events randomly distributed on a road network, and since their study is carried out once the event has occurred, we can consider that an underlying point process governs them on the linear network.

We are interested in finding the set of streets where serious traffic accidents are most likely to occur in a network of roads in a city. The collection of such streets must be formed mainly by a point process of high density that has a different intensity from the set of locations where serious traffic accidents are less likely to occur and which should occur over the whole street network of the city. The point pattern that occurs on the complete street network has a low density and makes sense in the context of severe traffic accidents to the extent that anywhere in the city, there may be streets that have a lower number of severe traffic accidents due to low traffic, impediments to high speeds, or other factors that decrease the risk of fatalities.

\subsection{Road traffic injuries and deaths in Bogota City}

Bogota is Colombia's capital and the largest and most populous city. In 2015, this city was estimated to have 2148541 vehicles \citep{SEB}. The simplified city road network used in this work consists of 199135 intersections, 243157 uninterrupted segments and a total length of 8218.90 km. We analyzed a database of 10979 traffic accidents that resulted in injuries or deaths in 2015 in the urban area of the city, published in the OpenData portal of Bogota Town Hall, together with the road network shapefile \citep{Moncada2018,SDM}. The city of Bogota is divided into eighteen zones called localities.  Identifying roads with high-density traffic injuries and deaths provides valuable information for decision-making in public policies that help prevent more of these unfortunate events. 

In the context of traffic accidents, we aim to employ our developed methodology in linear networks that allow us to distinguish highly congested zones in the sense of traffic accidents from those with lower intensity. The dataset consists entirely of traffic accidents and thus the classification of the traffic accidents into two distinct groups based on intensity: the most accidents rate and less accidents rate zones. By applying this innovative approach to separate patterns with higher traffic accident density from those with lower density, we can effectively identify and prioritize areas requiring additional attention and intervention regarding accident prevention and safety measures.

\begin{figure}[H]
\centering
\subfloat[Estimated intensity.]{\includegraphics[width=0.46\textwidth]{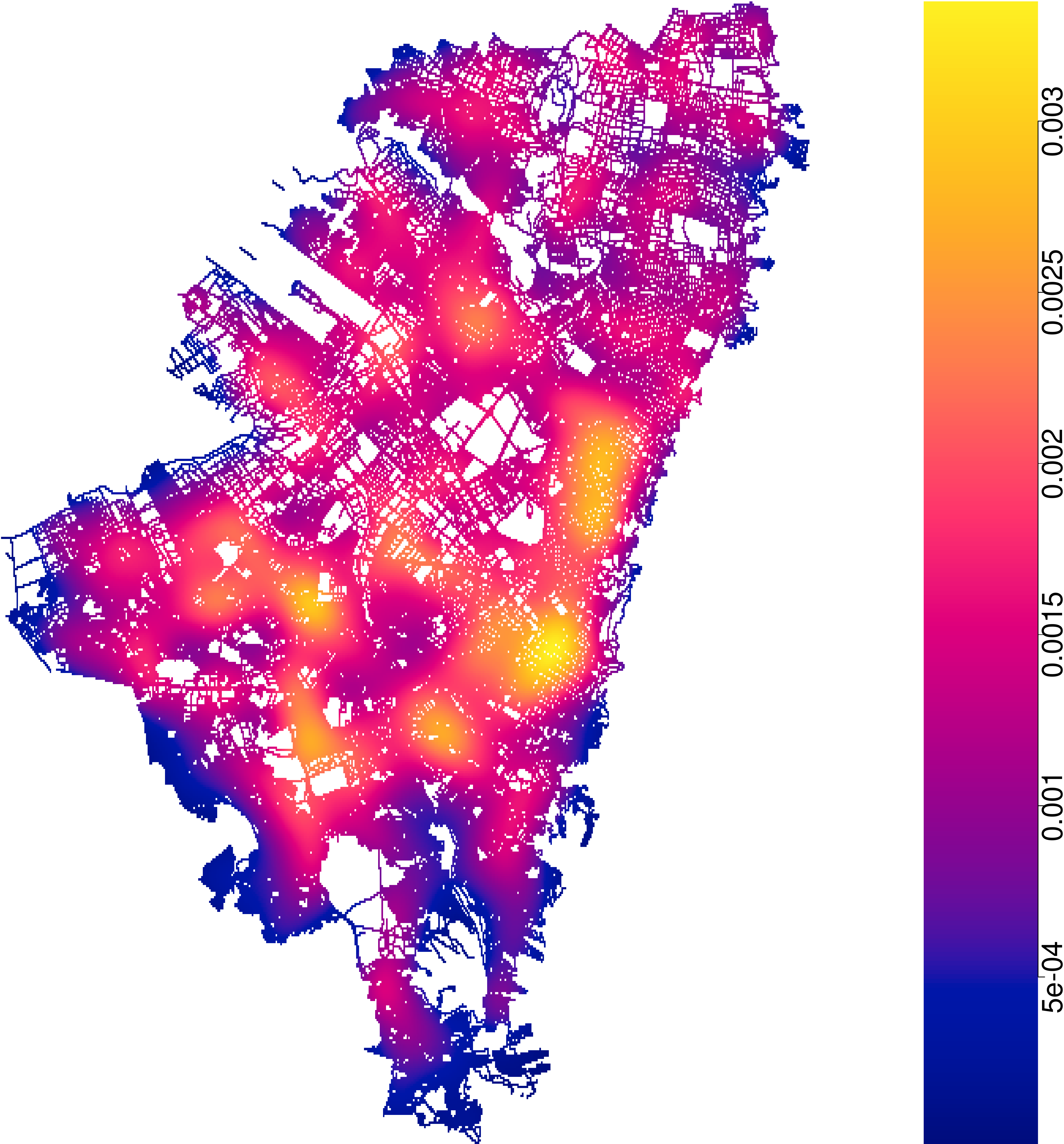}\label{fig:Intensity_Bog_600}}
\quad
\subfloat[Classification of the traffic accidents.]{\includegraphics[width=0.4\textwidth]{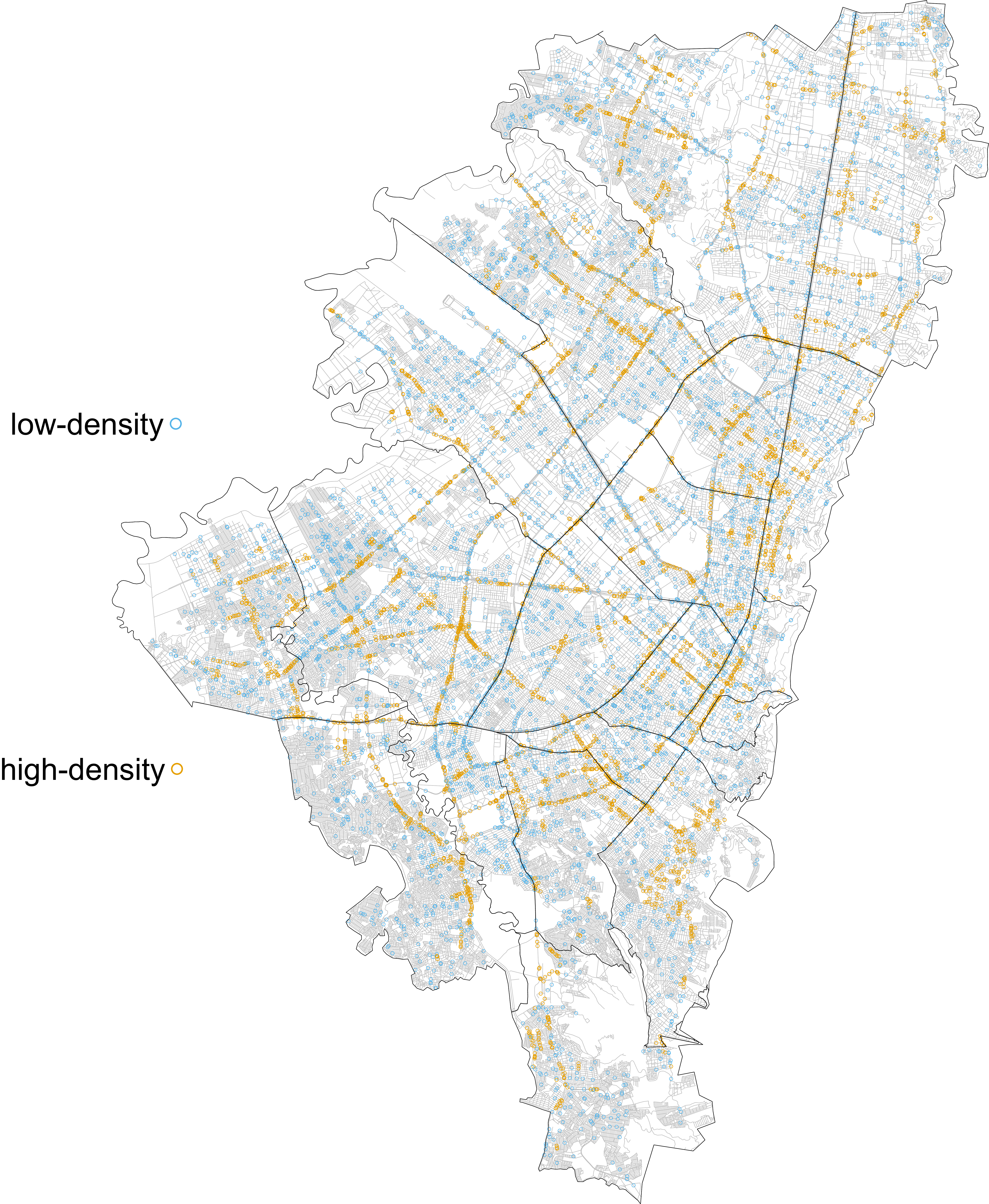}\label{fig:bogota_results}}
\caption{Traffic accidents on the road network of Bogota in which at least one person was injured or death recorded in the year 2015. \textit{Left:} Estimated non-parametric kernel-based intensity with bandwidth $\sigma=600$ meters. \textit{Right:} Classification of the traffic accidents in the localities of Bogota.}
\end{figure}

The application of the proposed classification procedure to analyze traffic accidents in Bogota directly using the whole network has a high computational cost, and the procedure does not yield results due to its size and complexity. To apply the proposed classification process to the Bogota network, it must be assumed that both overlapping point patterns are Poisson processes, which implies that the occurrence of events is independent in the disjoint sub-networks of this network. Therefore, we applied the classification procedure independently in each of the sub-networks of the eighteen localities of Bogota; then we consider the union of all these point patterns classified as the point pattern classified of traffic accidents in the whole network of Bogota as shown in Figure \ref{fig:bogota_results}. 

We estimate the non-parametric kernel-based intensity using a two-dimensional Gaussian kernel with a bandwidth of $\sigma=600$ meters \citep{RDMMGMB19} shown in Figure~\ref{fig:Intensity_Bog_600}. The bandwidth used was found empirically, starting from Scott's rule as a reference and then increasing it accordingly so that the hot spots would be displayed on the intensity plot. The kernel estimate of accident intensity in Bogota shows that most traffic accidents occur in different sectors around the city centre. However, a large part of the road network has intermediate intensity values, making it difficult to decide whether it is a road with a high accident rate. In this way, it would be possible to delimit some sub-networks where the probability of occurrence of new traffic accidents is high, but the roads with intermediate values make it difficult to decide on the delimitation of the sub-networks that must be intervened to avoid new traffic accidents. 

The classification of the traffic accidents in the city of Bogota during 2015 shown in Figure~\ref{fig:bogota_results} is depicted as a multitype point pattern for a dichotomous mark, in which it is possible to identify the road segments with a high density of traffic injuries and deaths in the city of Bogota. Therefore, the classification procedure can help to decide whether the road segments with intermediate intensity values are significantly relevant to delimit the sub-network and use this as a companion tool of the intensity to study the behaviour of traffic accidents in Bogota. 

In detail, the estimated intensity of the traffic injuries and death accidents in Bogota has a couple of large hot spots in the central-eastern zone of the city. This zone includes the downtown, the Palace of Justice and the national capitol, among other important places, surrounded by government offices, museums, churches and some historic buildings where many vehicles generally move. Additionally, an important avenue crosses the entire city from south to north and has access to many other avenues that pass through this zone. Other relevant hot spots in the city are located in the north zone, where the city’s financial, cultural, and recreational activity places are concentrated. The centre-west is another zone with a high probability of traffic accidents where large factories, parks, sports centres, administrative offices, and the international airport are located, which means that the roads that cross this zone are generally crowded with vehicles. Based on the estimated intensity and considering the classification results, it is possible to appreciate more explicitness of the road segments with high-density accidents in the hot spot zone. In summary, while Figure~\ref{fig:Intensity_Bog_600} shows the zones with the highest intensity values, the output in Figure~\ref{fig:bogota_results} helps to the delimitation of the sub-networks of higher traffic accidents and, in this way, to define which roads must be intervened with the highest priority to prevent traffic accidents. 

\subsection{Estimation of road traffic injuries and deaths in Medellin}
\begin{figure}[H]
\centering
\subfloat[Estimated intensity.]{\includegraphics[width=0.46\textwidth]{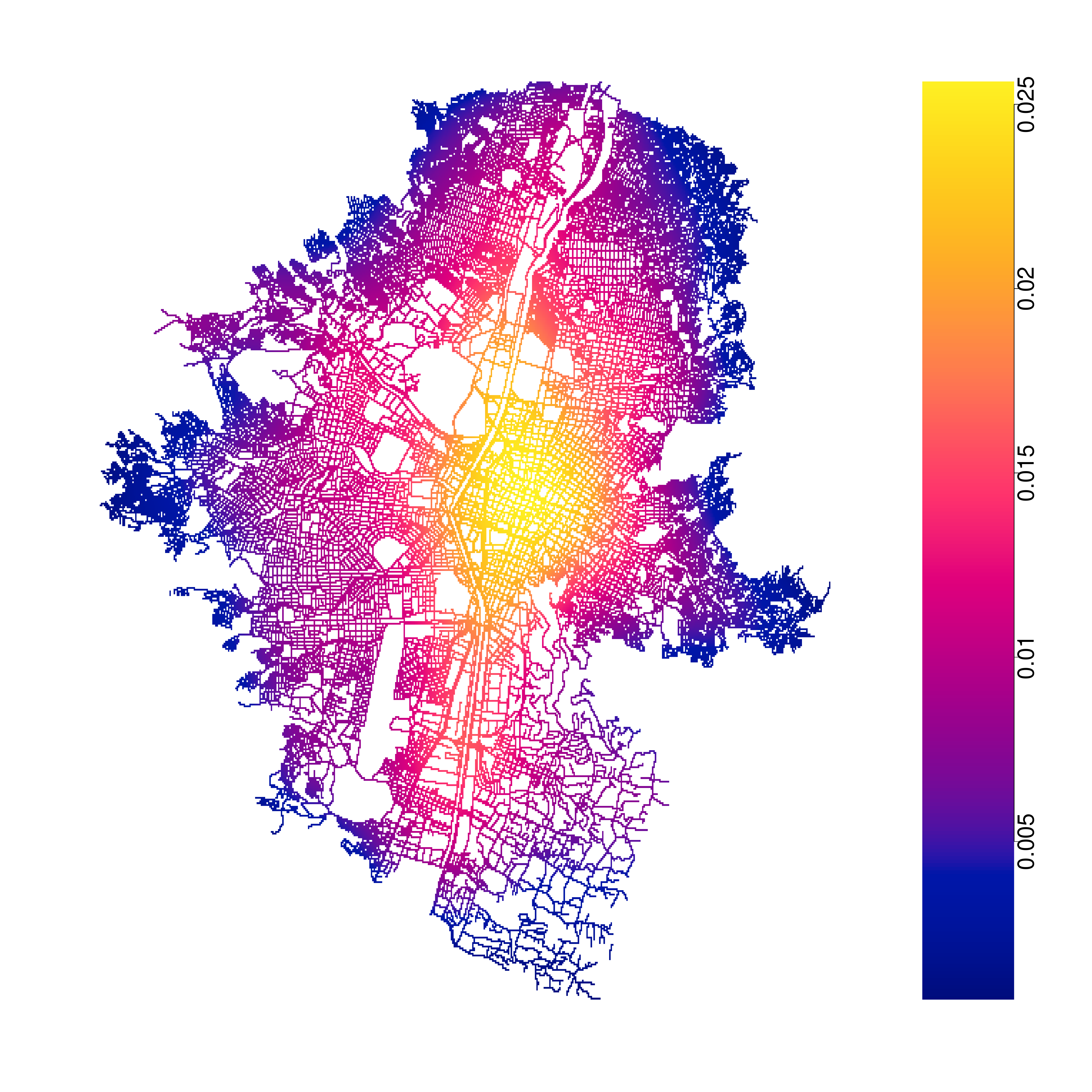}\label{fig:Intensity_Med_700}}
\quad
\subfloat[Classification of the traffic accidents.]{\includegraphics[width=0.4\textwidth]{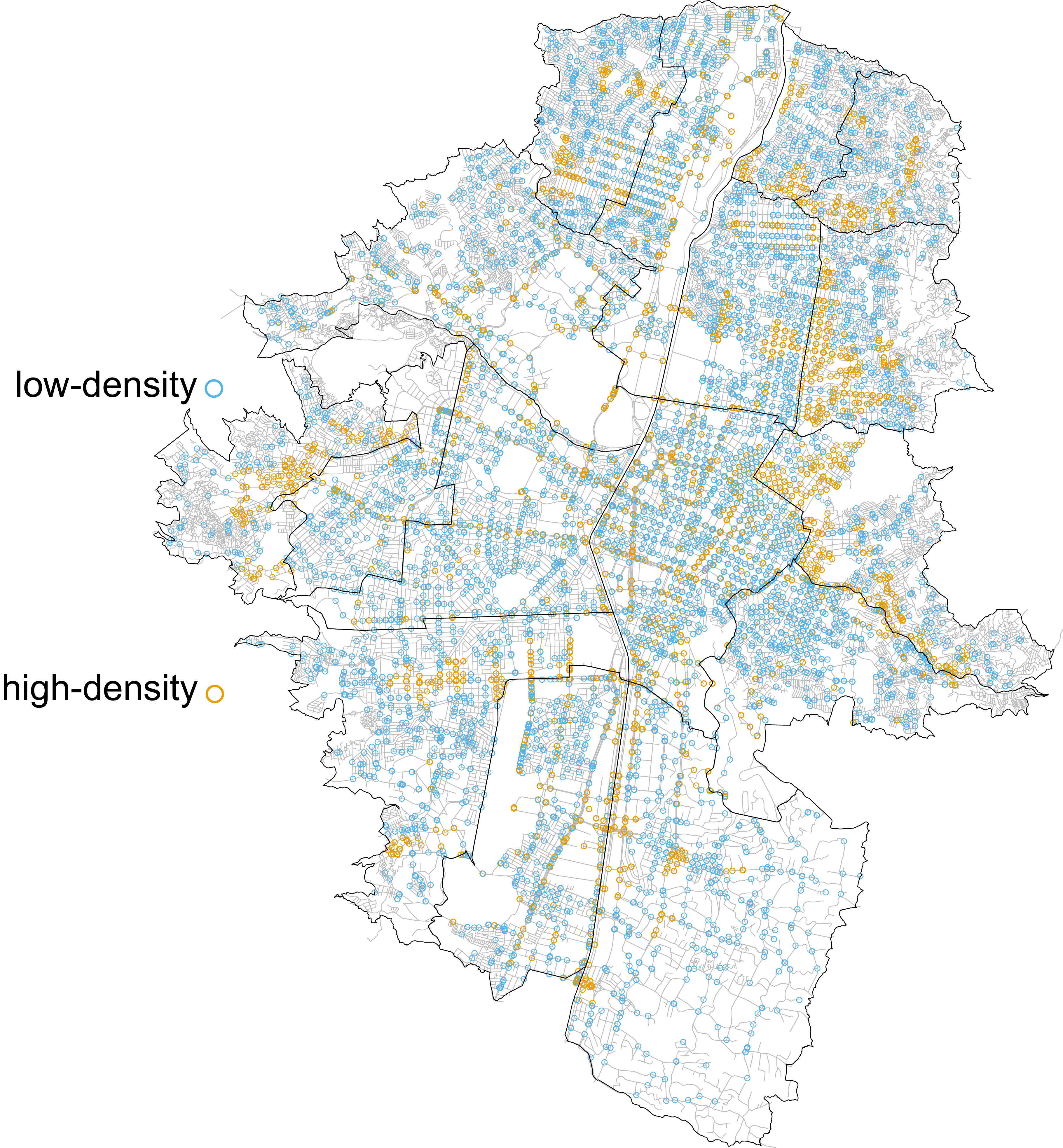}\label{fig:medellin_results}}
\caption{Traffic accidents on the road network of Medellin in which at least one person was injured or death recorded in the year 2019. \textit{Left:} Estimated non-parametric kernel-based intensity with bandwidth $\sigma=700$ meters. \textit{Right:} Classification of the traffic accidents in the communes of Medellin.}
\end{figure}

Medellin is the second most important city in Colombia after Bogota. In 2019 it was estimated that this city had 1756893 vehicles \citep{MCV}. The city’s road network consider in this study consists of 110533 intersections, 115939 uninterrupted segments and a total length of 2306.48 km. We analysed a database of 22743 traffic accidents that resulted in injuries or deaths in 2019 in the city, also published in the OpenData portal of Medellin Town Hall, and the road network shapefile \citep{SMM}. The city of Medellin is divided into sixteen zones called communes. Analogously as we did with Bogota, in Figure~\ref{fig:medellin_results}, we can see that we apply the classification procedure within each of the communes and then unify the classified point patterns to form the classified point pattern in the whole city of Medellin.

Figure~\ref{fig:Intensity_Med_700} shows the non-parametric kernel-based intensity estimated with bandwidth $\sigma=700$ meters. The intensity of the traffic injuries and death accidents in Medellin has a single large hot spot in the central-eastern zone which is the commercial and financial centre of the city. This zone includes the downtown, a large part of the city’s commerce, residences, factories, administrative offices, museums, churches, and historic buildings. Additionally, four important avenues cross the hot spot, and many vehicles circulate daily in the city using these roads. Based on the intensity in Figure~\ref{fig:Intensity_Med_700} and the result of the classification procedure of the traffic injuries and death accidents in Medellin in 2019 shown in Figure~\ref{fig:medellin_results}, it is possible to differentiate between the road segments with the highest number of accidents in the hot spots. Therefore,  the combination of the information provided by these two exploratory tools makes it more feasible to interpret the results in geometric spaces that are not intuitive.

\section{Conclusions}\label{sec:conclusions}
In this paper, we consider the problem of estimating features in a point pattern on a linear network in the presence of clutter. The proposed classification method allows for the distinction between clutter and feature in complex geometries and the parameters estimated by an $EM$ algorithm. Furthermore, it is possible to automatically select the number of nearest neighbours to consider in the statistical analysis through the segmented regression model. The methodology of classification can be applied without assumptions about the feature shapes, and it relies on the assumption that the feature and clutter are superimposed Poisson processes. 

The simulation study highlights the performance and accuracy of the proposed procedure in terms of classification rates. The classification procedure applied to the real point pattern dataset of traffic accidents occurring on the road network of a city, together with the estimation of intensity, can be a helpful tool in decision-making when delimiting the hot-spot roads that have priority to be intervened to prevent these accidents.

Given the growing interest in network data, we believe that the proposed approach could be used in many different contexts of application.
Future work may consider modifying the EM algorithm to estimate the intensities of more than two groups of features or clutter simulated over the same linear network, finding features with inhomogeneous intensity in the presence of clutter or extending this methodology to the spatio-temporal case on linear networks. 

\subsection*{Acknowledgments} 
The research work of Juan F. D\'iaz-Sep\'ulveda and Francisco J. Rodr\'guez-Cort\'es has been partially supported by Universidad Nacional de Colombia, HERMES projects, Grant/Award Number: 56470. The research work of Giada Adelfio and Nicoletta D'Angelo has been supported by ``FFR 2023 - Università degli Studi di Palermo - Giada Adelfio”, ``FFR 2023 - Università degli Studi di Palermo - Nicoletta D'Angelo”,  and by the PNRR project, grant agreement No PE0000018 - GRINS - Growing Resilient, Inclusive and Sustainable.

\bibliographystyle{apalike}
\bibliography{example}
\end{document}